\documentclass[twoside]{article}
\usepackage{qic,epsfig}

\usepackage{latexsym}
\usepackage{subfigure}
\usepackage{multirow}

\usepackage{url}

\begin{document}
\setlength{\textheight}{8.0truein}    %FOR 2ND PAGE ONWARDS

\runninghead{An $\Theta(\sqrt{n})$-depth Quantum Adder on a 2D NTC Quantum Computer Architecture}
            {Byung-Soo Choi and Rodney Van Meter}

\normalsize\textlineskip
\thispagestyle{empty}
\setcounter{page}{1}

%\copyrightheading{Vol.}{No.}{Year}{Page Nos.}
\copyrightheading{0}{0}{2010}{000--000}

\vspace*{0.88truein}

\alphfootnote

\fpage{1}

%%%%%%%%%%%%%%%%%%%%%
%Put in titiles here
%%%%%%%%%%%%%%%%%%%%%

\centerline{\bf AN $\Theta(\sqrt{n})$-DEPTH QUANTUM ADDER ON}
\vspace*{0.035truein}
\centerline{\bf THE 2D NTC QUANTUM COMPUTER ARCHITECTURE \footnote{A two-page short abstract was presented at AQIS 2010. This version includes all details of design and analysis of the proposed adder.}
}
\vspace*{0.37truein}

\centerline{\footnotesize BYUNG-SOO CHOI\footnote{Corresponding Author,\,\, bschoi3@gmail.com}}
\vspace*{0.015truein}
\centerline{\footnotesize\it Center for Quantum Information Processing}
\baselineskip=10pt
\centerline{\footnotesize\it Department of Electrical and Computer Engineering}
\baselineskip=10pt
\centerline{\footnotesize\it University of Seoul}
\baselineskip=10pt
\centerline{\footnotesize\it 13 Siripdae-gil (90 jeonnong-dong), Dongdaemun-gu, Seoul, 130-743, Republic of Korea}
\vspace*{10pt}

\centerline{\footnotesize RODNEY VAN{ }METER\footnote{rdv@sfc.wide.ad.jp}}
\vspace*{0.015truein}
\centerline{\footnotesize\it Faculty of Environment and Information Studies}
\baselineskip=10pt
\centerline{\footnotesize\it Keio University}
\baselineskip=10pt
\centerline{\footnotesize\it 5322 Endo, Fujisawa, Kanagawa, 252-8520, Japan}
\vspace*{10pt}

\vspace*{0.225truein}
\publisher{(\today)}{(revised date)}

\vspace*{0.21truein}

%% \abstracts{first paragraph}{second paragraph}{third paragraph}
%% If there is only one paragraph, just keep the second and third empty
%% like the following one
\abstracts
{
In this work, we propose an adder for the 2D NTC architecture, designed to match the architectural constraints of many quantum computing technologies. The chosen architecture allows the layout of logical qubits in two dimensions and the concurrent execution of one- and two-qubit gates with nearest-neighbor interaction only. The proposed adder works in three phases. In the first phase, the first column generates the summation output and the other columns do the carry-lookahead operations. In the second phase, these intermediate values are propagated from column to column, preparing for computation of the final carry for each register position. In the last phase, each column, except the first one, generates the summation output using this column-level carry. The depth and the number of qubits of the proposed adder are $\Theta(\sqrt{n})$ and $O(n)$, respectively. The proposed adder executes faster than the adders designed for the 1D NTC architecture when the length of the input registers $n$ is larger than 58.
}
{}{}

\vspace*{10pt}

\vspace*{3pt}
\communicate{to be filled by the Editorial}

\vspace*{1pt}\textlineskip    %) USE THIS MEASUREMENT WHEN THERE IS
   %) A SECTION HEADING
%\vspace*{-0.5pt}
%\noindent
%%%%%%%%%%%%%%%%%%%%%%%%%%%%%%%%
%put the text of the paper here
%%%%%%%%%%%%%%%%%%%%%%%%%%%%%%%%
\section{Introduction}

Quantum computers have been proposed to exploit the exotic properties of quantum mechanics for information processing. Among many potential uses, two quantum algorithms have received the bulk of the attention. One is Shor's large number factoring algorithm \cite{shor-factoring}, and the other is Grover's unstructured database search algorithm \cite{Grover-search}, though there has also been much progress recently on other algorithms \cite{mosca-2008,Bacon-vanDam-2010-Algo-Review,Brown:1263035}. Quantum algorithms are often shown to be more efficient than classical ones by analyzing the number of queries to an oracle. However, for a more exact performance analysis, we need to analyze the quantum algorithms in terms of the detailed quantum circuits necessary to implement them. Among many circuits, as in classical computation, a core set of subroutines whose behavior will strongly impact the performance of the overall algorithm is arithmetic, hence we focus on the adder in this work.

Numerous quantum addition circuits have been proposed using abstract models of the computer itself. The basic elementary quantum arithmetic operations including addition have been proposed by Vedral et al. \cite{VBE-adder} and Beckman et al. \cite{BCDP-adder}, following seminal work on elementary reversible full- and half-adders by Fredkin and Toffoli \cite{fredkin82:_conserv_logic}, and Feynman \cite{feynman:lect-computation}. Glassner proposed an one-qubit full adder \cite{DBLP:journals/cga/Glassner01d}. Subsequently, Cheng and Tseng proposed an \emph{n}-qubit full adder and subtractor based on the work of Glassner \cite{Cheng-Tseng-adder}. Reducing the space requirements for those earlier adders \cite{VBE-adder}, Cuccaro et al. proposed a linear-depth ripple-carry adder with only a single ancillary qubit \cite{CDKM-adder}. Meanwhile Draper proposed a transform adder based on the quantum Fourier transform \cite{Draper-QAddition}. Draper et al. proposed a fast quantum carry-lookahead adder \cite{draper-QCLA}. Takahashi and Kunihiro have shown that addition can actually be performed with no ancillae, at the expense of a deeper circuit \cite{takahashi-2005}.

Incorporating the behavior of these circuits, we can estimate the overall quantum speedup more accurately than simply addressing the issue at the query-level, and confirm again that the quantum speedup is very high. However, it is not possible to determine the exact performance gain unless the practical issues of architecture are considered; both the constant factors and the leading order of both the computational complexity and minimum execution time (or circuit depth) depend on the assumed underlying machine. Hence we have to consider many issues such as error correction, communication, gate, and qubit technologies \cite{DBLP:series/synthesis/2006Metodi}. For example, Maslov et. al \cite{Maslov-QCircuitPlacement} pointed out the importance of the problem of placing circuit variables on the underlying qubit layout. Unfortunately, it is impossible to consider all practical issues at the same time. To avoid this problem, we usually define a practical quantum computer architecture incorporating as many practical constraints as possible. For many quantum computer architectures, the 2D NTC architecture is a reasonable model capturing the key factors that impact performance. NTC allows \emph{N}earest-neighbor interactions, \emph{T}wo-qubit quantum gates, and \emph{C}oncurrent executions of gates \cite{VBE-Improved}. An example of a potentially scalable architecture with the nearest-neighbor constraint is that of Kielpinski et. al \cite{wineland_scaling}. Barenco et al. \cite{Barenco-ElementaryGates} showed one way to decompose a given quantum circuit into two-qubit gates. Steane \cite{steane-reliable-quantum-computing} investigated the necessity  of concurrent execution for error correction and fault-tolerance; concurrency is also required at the application level for high performance. The 2D allows a single qubit to interact with four neighboring qubits. With more neighboring qubits than the 1D case, the 2D layout should show higher performance, thanks to reduced distance between many pairs of qubits and the potential for more concurrent movement of qubits. Likewise, a 3D layout should show higher performance than the 2D case, but the complexity of fabricating and controlling qubits in three dimensions likely makes it impractical. Therefore, we believe that the 2D layout is the most reasonable choice at the middle level of performance and control overhead. Thus, it would be interesting to understand the quantum speedup in this context.

Surprisingly, as far as we know, there is no quantum addition circuit designed specifically for the 2D NTC architecture. Hence we have to design a quantum addition circuit for the 2D NTC architecture and estimate the performance gain. Based on this, our contributions are as follows.

\begin{itemize}
\item   {\textbf{Propose a quantum adder on the 2D NTC architecture.}} \newline
First, we lay out the qubits in a $\sqrt{n}\times\sqrt{n}$ array where $n$ is the input size, in qubits. Based on this layout, we propose a three-phase quantum addition algorithm. In the first phase, the first column does a ripple-carry addition and the other columns do carry-lookahead operations. In the second phase, the column-level carry is propagated in ripple fashion  between the columns. In the last phase, each column transports its column-level carry input into the cells to generate the final summation value.

\item   {\textbf{Analyze the proposed adder.}} \newline
We decompose the necessary quantum circuit blocks using only one- and two-qubit gates. Next, we add \textbf{SWAP} operations necessary to transport qubits in order to satisfy the NTC constraint. We found that the depth of the proposed adder is $150\sqrt{n}-90$ in terms of one- and two-qubit gates. Asymptotically, the depth is $\Theta(\sqrt{n})$ meeting the depth lower bound we established in earlier work \cite{choi-2008}. To execute many  quantum gates in parallel, the proposed adder utilizes many working qubits as $2n-\sqrt{n}$ qubits.

Since the 2D NTC layout generalizes the 1D NTC architecture, the adders designed for the 1D NTC architecture can also be implemented on the 2D NTC architecture without modification. After reevaluating the depth of the adders for the 1D NTC architecture, we find that our new 2D adder works faster when $n \geq 58$.

\end{itemize}

This paper is organized as follows. We explain the addition algorithm, and qubit and circuit layouts for the 2D NTC architecture in Section \ref{sec:2d-adder}. The temporal and spatial resources are analyzed in Section \ref{sec:analysis}. Finally, we conclude this work and point out some problems in Section \ref{sec:conclusion}.

\section{Adder on the 2D NTC Structure}
\label{sec:2d-adder}

In this section, we first explain how the qubits are laid out on the 2D structure firstly. Second, we explain an addition algorithm based on a slight modification of carry-lookahead addition. Third, we discuss how the addition algorithm is mapped with the circuit blocks. Finally, we show how the ancillae qubits can be initialized.

\subsection{Qubit Layout}

\begin{figure}[t]
\centerline{\epsfig{file=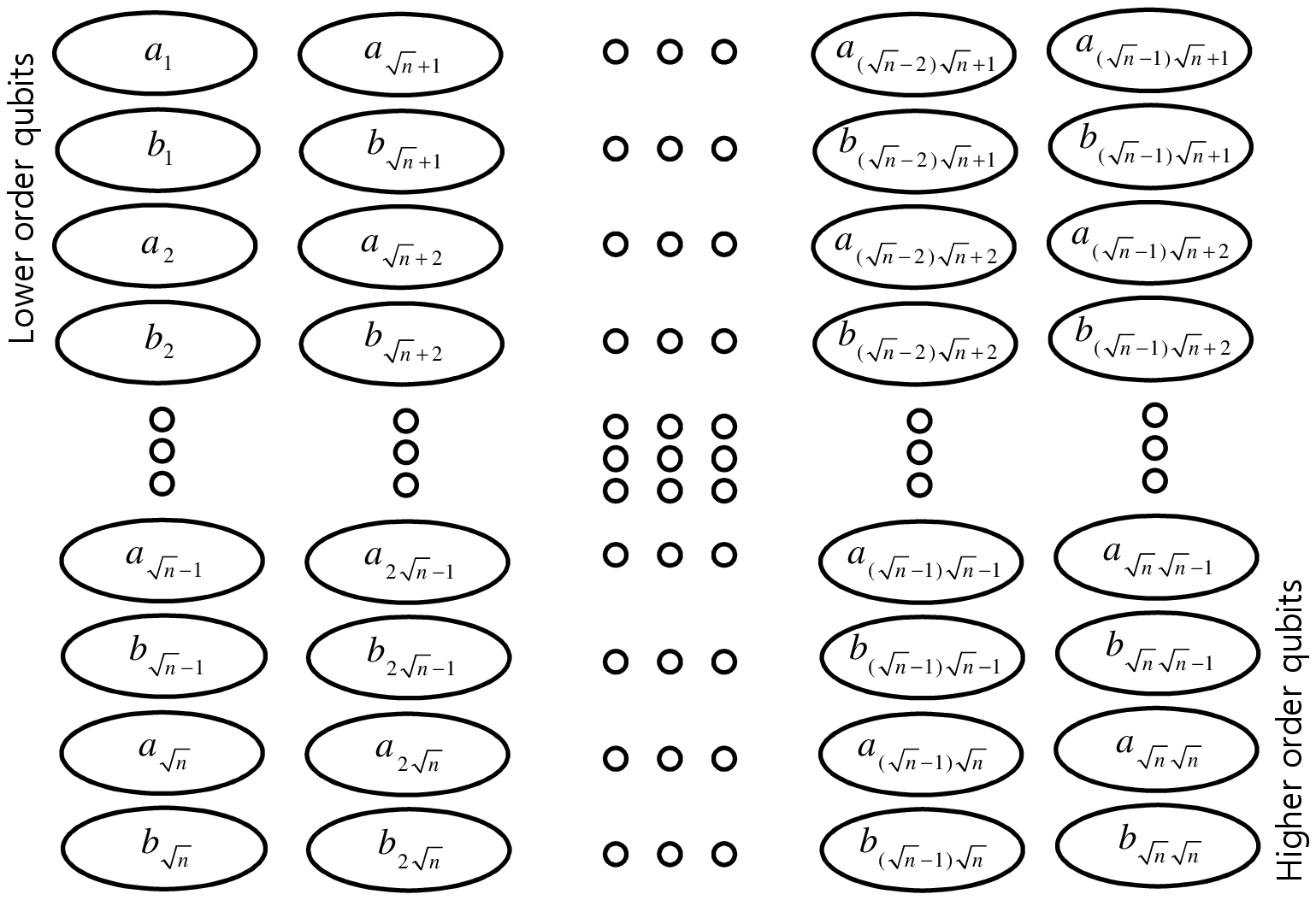, scale=0.6}} %100 percent
\vspace*{13pt}
\fcaption{\label{2d-qubit-layout} Layout of Input Qubits for a 2D NTC adder. \\
Two inputs are $A=a_n \cdot 2^{n-1}+a_{n-1} \cdot 2^{n-2}+\cdots+a_1$ and $B=b_n \cdot 2^{n-1}+b_{n-1} \cdot 2^{n-2}+\cdots+b_1$. $i$-th qubit is located at $(k,j)$ position where $k=\lceil i/\sqrt{n} \rceil$ and $j=i-(k-1)\sqrt{n}$. Ancillae qubits are not shown for simplicity.
}
\end{figure}

On the 2D NTC structure, we can lay out the qubits as shown in Figure \ref{2d-qubit-layout}. In the figure, the two input registers are $A=a_n \cdot 2^{n-1}+a_{n-1} \cdot 2^{n-2}+\cdots+a_1$ and $B=b_n \cdot 2^{n-1}+b_{n-1} \cdot 2^{n-2}+\cdots+b_1$. As shown in the figure, the two inputs $a_i$ and $b_i$ are interleaved where $1 \leq i \leq n$. The number of rows and columns are $2\sqrt{n}$ and $\sqrt{n}$, respectively. Two inputs $a_i$ and $b_i$ are located at a ($k$-th column, $j$-th row) cell where $k=\lceil i / \sqrt{n} \rceil$ and $j=i-(k-1)\sqrt{n}$. The figure shows only the input qubits for clarity. For simplicity, we assume without loss of generality that $\sqrt{n}$ is an integer.

\subsection{Adapting Carry-Lookahead Addition to Limited Interaction Distance}

To set the stage for the later arithmetic discussions, let us first explain the ripple for two $n$-qubit input registers, $a$ and $b$. Since the summation value for the $i$-th position $s_i$ is generated by $a_i \oplus b_i \oplus c_i$, where $a_i$ and $b_i$ are the $i$-th qubits in the input registers, and $c_i$ is the carry input from the summation of the $(i-1)$-th position, the time complexity of the addition depends on how fast the carry information can be transported between the bit positions.

The simplest circuit is the ripple carry adder, which propagates the carry information stepwise from position to position. The carry output for the $(i+1)$-th position, $c_{i+1}$, should be one if a majority of the bits $a_i$, $b_i$, and $c_i$ are one, and zero otherwise; it is generated by $a_i \cdot b_i \oplus a_i \cdot c_i \oplus b_i \cdot c_i$. Therefore, the final summation value $s_n$ is generated only after $n$ ripple carry time steps.

To reduce this time, a carry-lookahead method was devised. In this method, two additional values are defined as follows:
\begin{eqnarray}
g_i &   =   &   a_i \cdot   b_i. \\ \newline
p_i &   =   &   a_i \oplus  b_i.
\end{eqnarray}
Implicitly, $g_i$ and $p_i$ determine whether this bit position \emph{generates} a carry out independent of the carry in, or \emph{propagates} its incoming carry to its output carry, respectively. Only one of these may be true, though both may be false (called carry \emph{kill}, though kill is not necessary in the actual circuit). The carry output for $(i+1)$-th position is generated as $c_i = g_i \oplus p_i \cdot c_{i-1}$. Therefore, if $g_i$ is one, $c_i$ has no dependence on $c_{i-1}$, and hence disconnects the carry chain. However, if $g_i$ is zero, $c_i$ is dependent on $c_{i-1}$. In the worst case, the longest chain is from $c_1$ to $c_n$. To decompose this long chain into sub-units, two variables $G[i,j]$ and $P[i,j]$ are also defined as follows.
\begin{eqnarray}
G[i,j]  &   =   &   g_j \oplus p_j \cdot G[i,j-1]. \\ \newline
P[i,j]  &   =   &   p_j \cdot  P[i,j-1].
\end{eqnarray}
$G[i,j]$ indicates whether an entire \emph{span} of the addition, from qubit $i$ to qubit $j$, generates a carry.  Similarly, $P[i,j]$ indicates whether the $[i,j]$ span propagates the carry from position $i$ all the way to position $j$. By calculating these values concurrently and progressively increasing the span of $G$ and $P$, the total time to create complete carry information for the entire register can be reduced to $O(\log n)$, provided that communication within the system is adequately fast.

Unfortunately, this carry-lookahead addition algorithm is defined assuming no limitation of interaction distance, and hence cannot be applied for the 2D NTC architecture without modification. In this work, we slightly modify the carry-lookahead, which consists of three phases as follows.

\subsubsection{Phase 1: Ripple Carry Addition on the First Column, and Carry-Lookahead on the Other Columns}

\begin{figure}[t]
\centerline{\epsfig{file=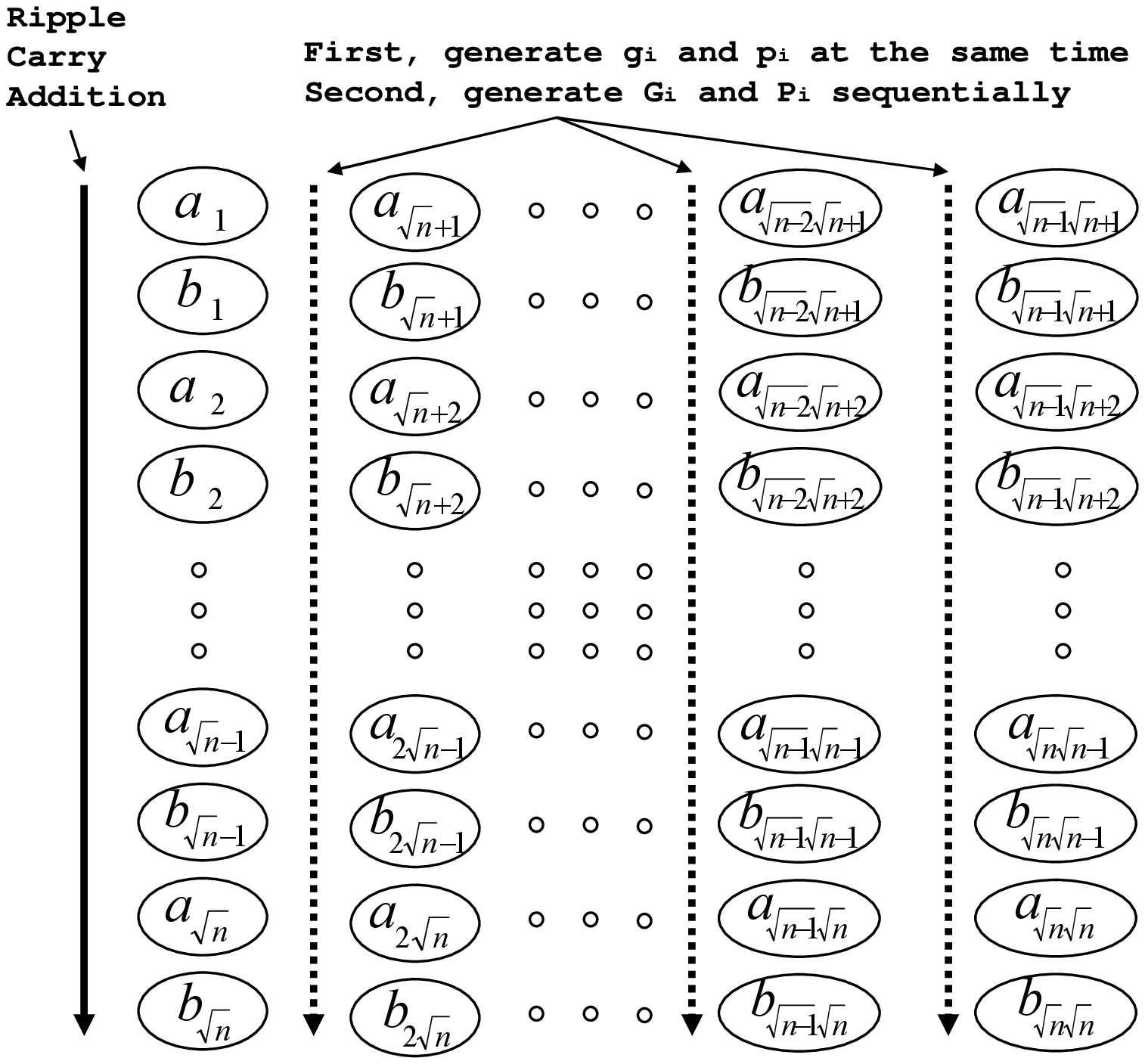,scale=0.5}} %100 percent
\vspace*{13pt}
\fcaption{\label{first-step}First phase. \newline
During this phase, the first column executes a ripple-carry adder. The other $k$-th column generates $g_{(k-1)\sqrt{n}+j}$ and $p_{(k-1)\sqrt{n}+j}$ concurrently, and then $G[(k-1)\sqrt{n}+1,(k-1)\sqrt{n}+j]$ and $P[(k-1)\sqrt{n}+1,(k-1)\sqrt{n}+j]$ sequentially.}
\end{figure}

As shown in Figure \ref{first-step}, the first column does the typical ripple carry addition. From the first position to the last position, each position generates a summation value and a carry output as follows.
\begin{eqnarray}
s_{i}           &   =   &   a_{i}\oplus b_{i} \oplus c_{i},  \label{eqn:summation}  \\
c_{i+1}         &   =   &   a_{i}\cdot b_{i} \oplus a_{i}\cdot c_{i} \oplus b_{i}\cdot c_{i},
\end{eqnarray}
where $c_1=0$. Since the carry output of the $i$-th position must be used as input for the next $(i+1)$-th position, there is an information dependency, hence this step takes about $O(\sqrt{n})$ time.

During this time, the other columns concurrently generate other necessary information for carry-lookahead operations. For example, the $k$-th column works as follows. First, each $(k,j)$ cell generates $g_{(k-1)\sqrt{n}+j}$ and $p_{(k-1)\sqrt{n}+j}$ \emph{concurrently},
\begin{eqnarray}
g_{(k-1)\sqrt{n}+j}   &   =   &   a_{(k-1)\sqrt{n}+j} \cdot b_{(k-1)\sqrt{n}+j}, \label{eqn:small_g_i}   \\
p_{(k-1)\sqrt{n}+j}   &   =   &   a_{(k-1)\sqrt{n}+j} \oplus b_{(k-1)\sqrt{n}+j}, \label{eqn:small_p_i}
\end{eqnarray}
where $1\leq j \leq \sqrt{n}$. After that, each $(k,j)$ cell generates $G[{(k-1)\sqrt{n}+1}, {(k-1)\sqrt{n}+j}]$ and $P[{(k-1)\sqrt{n}+1}, {(k-1)\sqrt{n}+j}]$ \emph{sequentially},
\begin{eqnarray}
G[(k-1)\sqrt{n}+1,(k-1)\sqrt{n}+j] & =  & g_{(k-1)\sqrt{n}+j} \oplus \label{eqn:large_g_i} \\ \nonumber
                                   &    & p_{(k-1)\sqrt{n}+j} \cdot G[(k-1)\sqrt{n}+1,(k-1)\sqrt{n}+j-1], \\
P[(k-1)\sqrt{n}+1,(k-1)\sqrt{n}+j] & =  & p_{(k-1)\sqrt{n}+j} \cdot P[(k-1)\sqrt{n}+1,(k-1)\sqrt{n}+j-1], \label{eqn:large_p_i}
\end{eqnarray}
where $G[(k-1)\sqrt{n}+1, (k-1)\sqrt{n}+1]=g_{(k-1)\sqrt{n}+1}$ and $P[(k-1)\sqrt{n}+1, (k-1)\sqrt{n}+1]=p_{(k-1)\sqrt{n}+1}$. The same process is applied for the other columns.

After this phase, the first column generates its final summation output and also the carry output $c_{\sqrt{n}+1}$. The other columns generate the column-level carry-lookahead information $G[(k-1)\sqrt{n}+1, k\sqrt{n}]$ and $P[(k-1)\sqrt{n}+1, k\sqrt{n}]$.

\subsubsection{Phase 2: Inter-Column Carry Propagation}
\begin{figure}[t]
\centerline{\epsfig{file=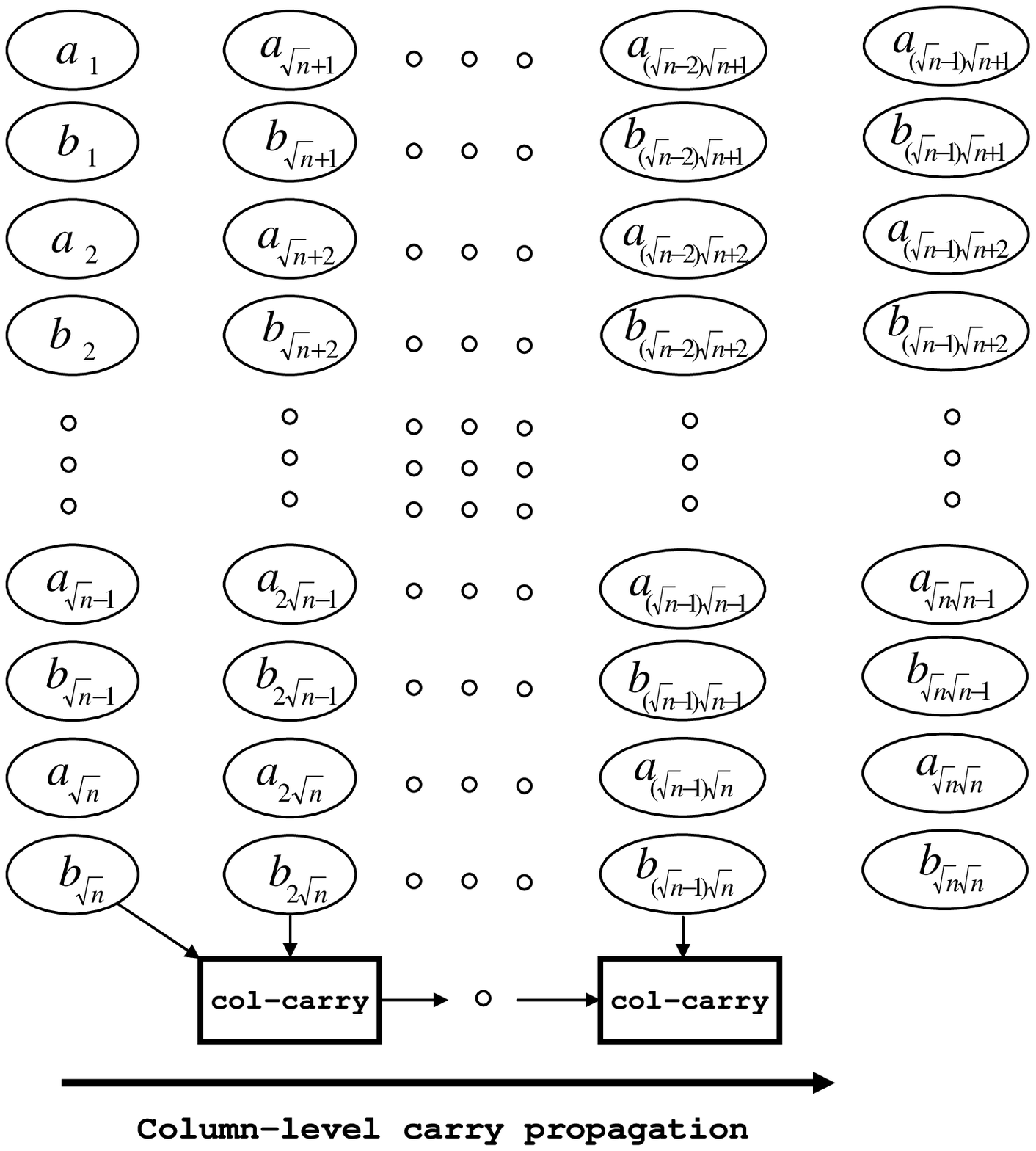,scale=0.5}} %100 percent
\vspace*{13pt}
\fcaption{\label{second-step}Second phase. \newline
The purpose of this phase is to generate column-level carry output for each column sequentially.}
\end{figure}

The final carry output of the first column, $c_{\sqrt{n}+1}$, is given as an initial input value for the column-level carry generation logic as shown in Figure \ref{second-step}. Each column, except the first, generates its column-level carry output as follows.
\begin{equation}
Column\_carry_k=c_{k\sqrt{n}+1}=G[(k-1)\sqrt{n}+1,k\sqrt{n}] \oplus c_{(k-1)\sqrt{n}+1} \cdot P[(k-1)\sqrt{n}+1,k\sqrt{n}]. \label{eqn:col_carry}
\end{equation}

\subsubsection{Phase 3: Carry Generation and Summation}

\begin{figure}[t]
\centerline{\epsfig{file=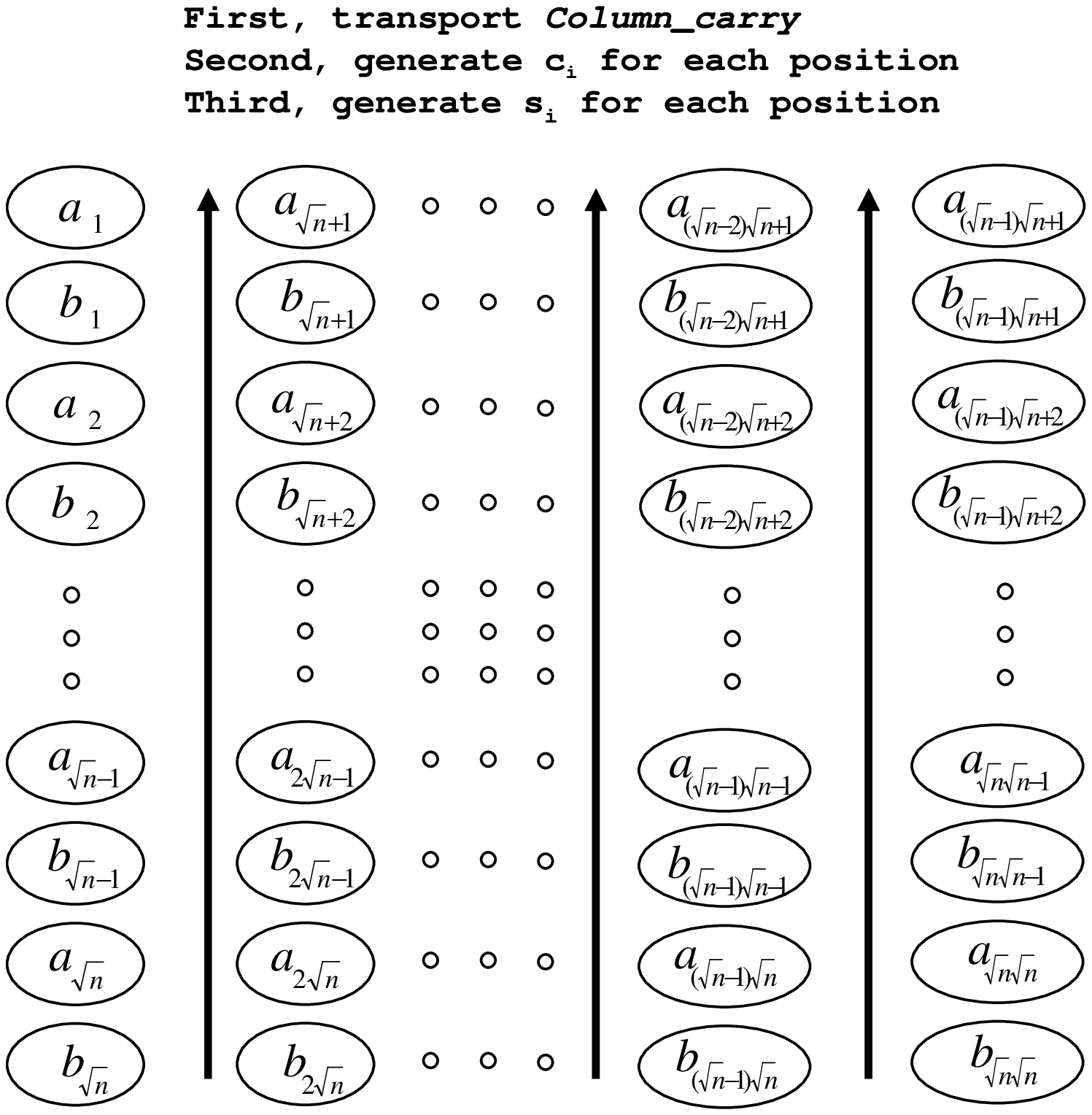,scale=0.5}} %100 percent
\vspace*{13pt}
\fcaption{\label{third-step}Third phase. \newline
Using the incoming carry for each column, all carry and sum are generated sequentially.}
\end{figure}

After the first phase, each $(k,j)$ cell has the carry-lookahead information $G[(k-1)\sqrt{n}+1,(k-1)\sqrt{n}+j]$ and $P[(k-1)\sqrt{n}+1,(k-1)\sqrt{n}+j]$. After the second phase, each column has the incoming column-level carry $c_{(k-1)\sqrt{n}+1}$. By propagating incoming column-level carry as shown in Figure \ref{third-step}, each $(k,j)$ cell can calculate its final carry input as
\begin{eqnarray}
c_i = c_{(k-1)\sqrt{n}+j}     &   =   &   G[(k-1)\sqrt{n}+1,(k-1)\sqrt{n}+j] \\ \nonumber
                        &       &   \oplus c_{(k-1)\sqrt{n}+1} \cdot P[(k-1)\sqrt{n}+1,(k-1)\sqrt{n}+j]. \label{eqn:carry_with_G_P}
\end{eqnarray}
After that, each cell can generate the final summation value as
\begin{equation}
s_i = s_{(k-1)\sqrt{n}+j}=a_{(k-1)\sqrt{n}+j} \oplus b_{(k-1)\sqrt{n}+j} \oplus c_{(k-1)\sqrt{n}+j}.
\end{equation}

\subsection{Circuit Layout}

\begin{figure}[t]
\centering
\subfigure[\label{first-step-circuit-flow-RCA}First Column]{\epsfig{file=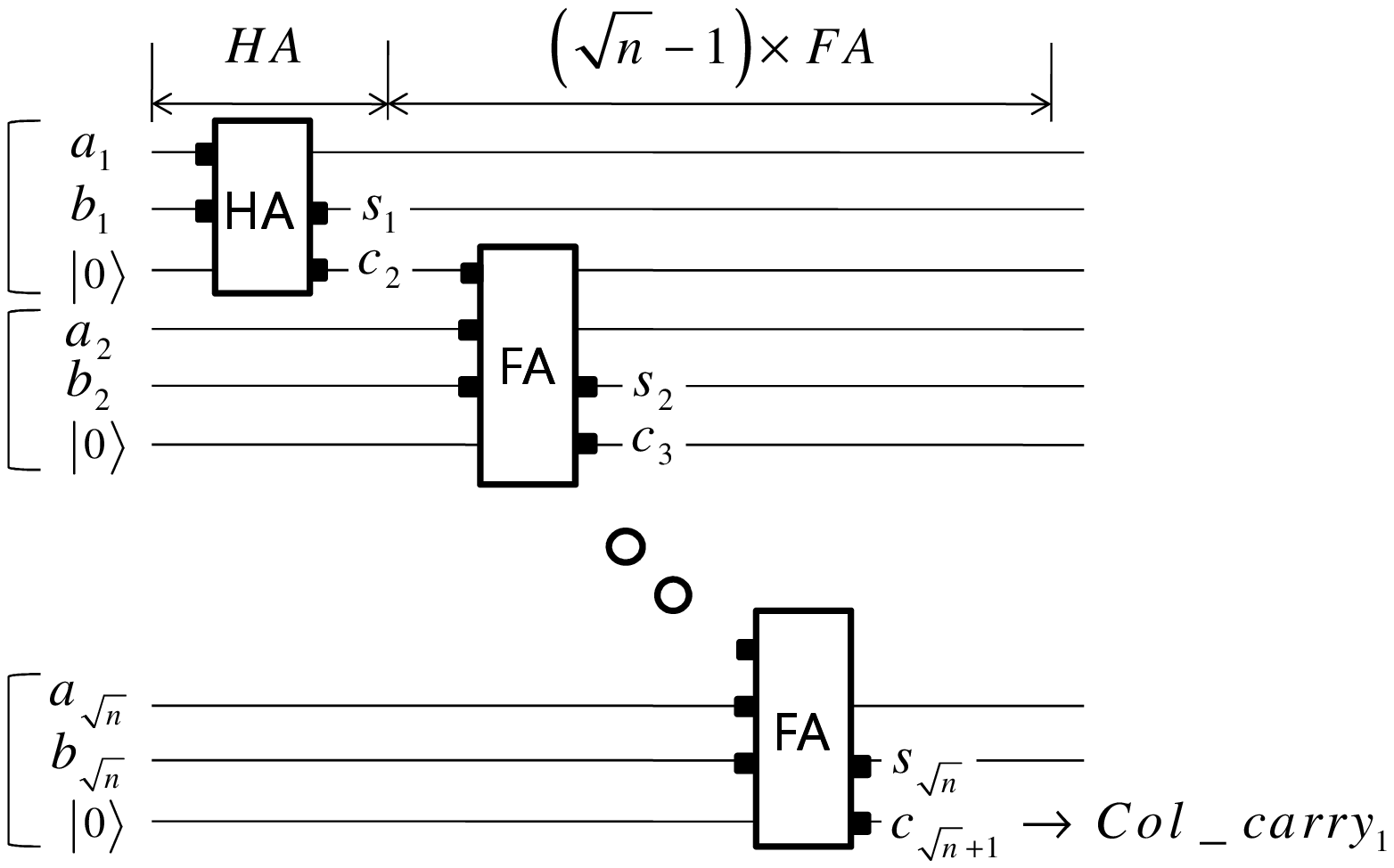,scale=0.35}}
\hspace*{10pt}
\subfigure[\label{first-step-circuit-flow-GP}Other Columns]{\epsfig{file=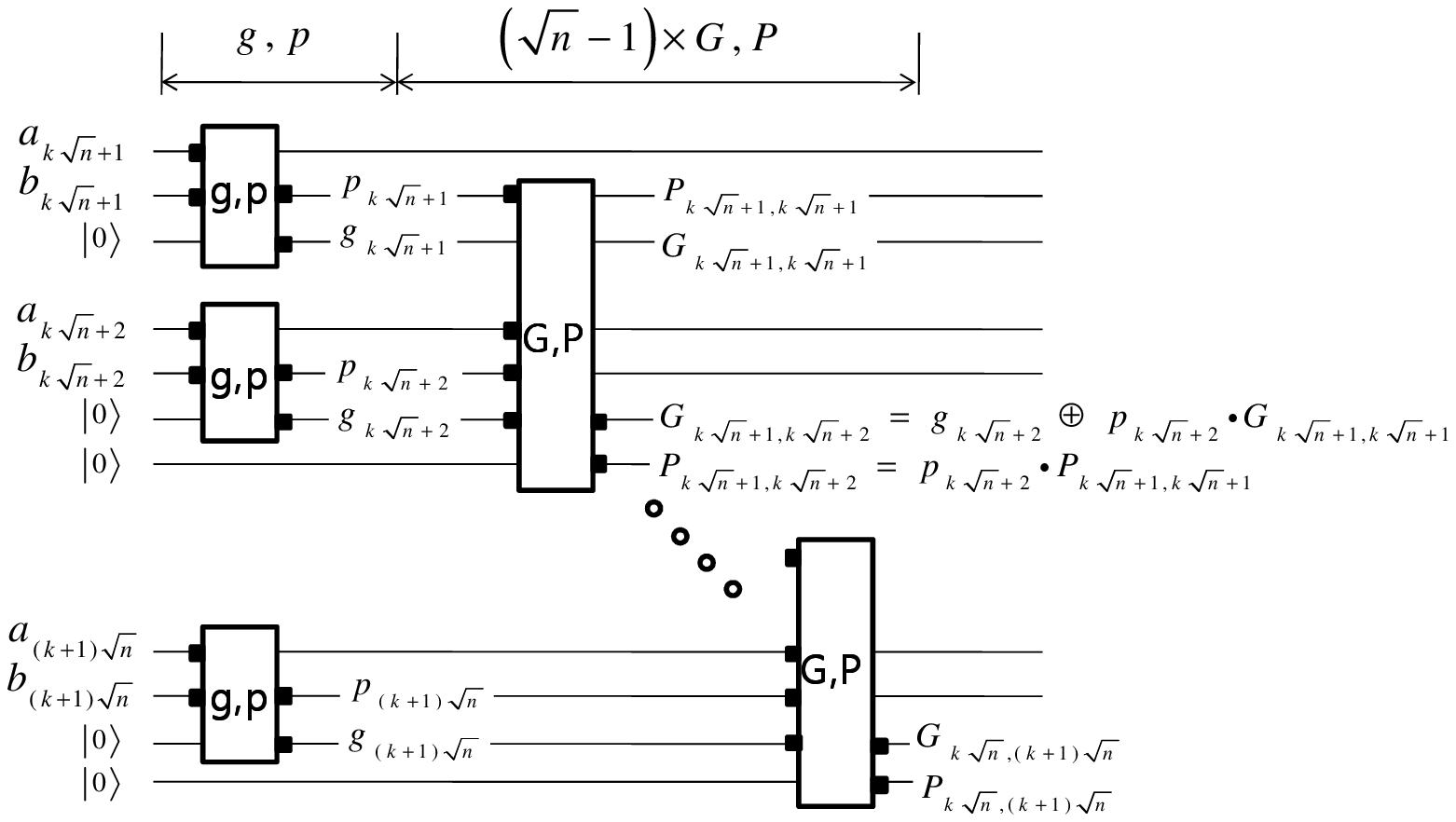,scale=0.45}}
%\vspace*{13pt}
\fcaption{Circuit flow for the first phase. \newline
Note \textbf{FA} and \textbf{HA} are the full adder and the half adder, respectively. $s$ and $c$ are initially $|0\rangle$, and $s_i$ and $c_i$ are summation and carry for each position, respectively.}
\end{figure}

In the first phase, the first column and the other columns use different circuit blocks. The circuit blocks for the first column are shown in Figure \ref{first-step-circuit-flow-RCA}. To do the ripple carry addition, a half-adder (\textbf{HA}) for the first position and $\sqrt{n}$-1 full-adders (\textbf{FA}) are used. The circuit blocks for the other columns are shown in Figure \ref{first-step-circuit-flow-GP}. As explained in the previous part, it generates first $g_{(k-1)\sqrt{n}+j}$ and $p_{(k-1)\sqrt{n}+j}$ \emph{concurrently} by using the \textbf{g,\,p} circuit blocks and then $G[(k-1)\sqrt{n}+1,(k-1)\sqrt{n}+j]$ and $P[(k-1)\sqrt{n}+1,(k-1)\sqrt{n}+j]$ \emph{sequentially} by using the \textbf{G,\,P} circuit blocks.

\begin{figure}[t]
\centerline{\epsfig{file=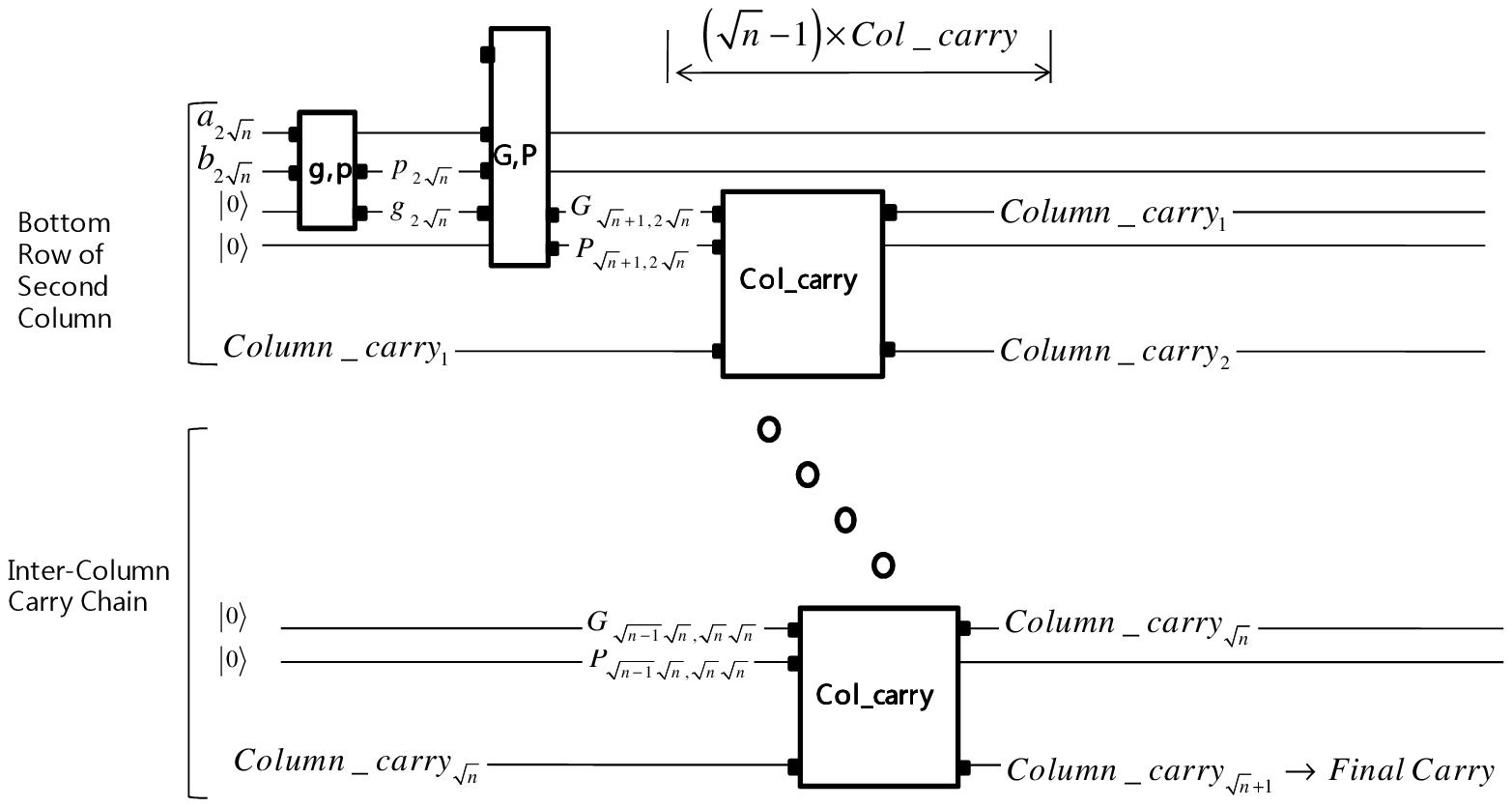,scale=0.8}} %100 percent
\vspace*{13pt}
\fcaption{\label{second-step-circuit-flow}Circuit flow for the second phase. \newline
\textbf{Col-carry} block generates a column-level carry output, which is used for the actual incoming carry value for the next (right) column.}
\end{figure}

The block-level circuit for the second phase is shown in Figure \ref{second-step-circuit-flow}. The circuit block \textbf{Col-carry} has three inputs: $G$ and $P$ from the corresponding column and $Column\_carry$ from the lower column.

\begin{figure}[t]
\centerline{\epsfig{file=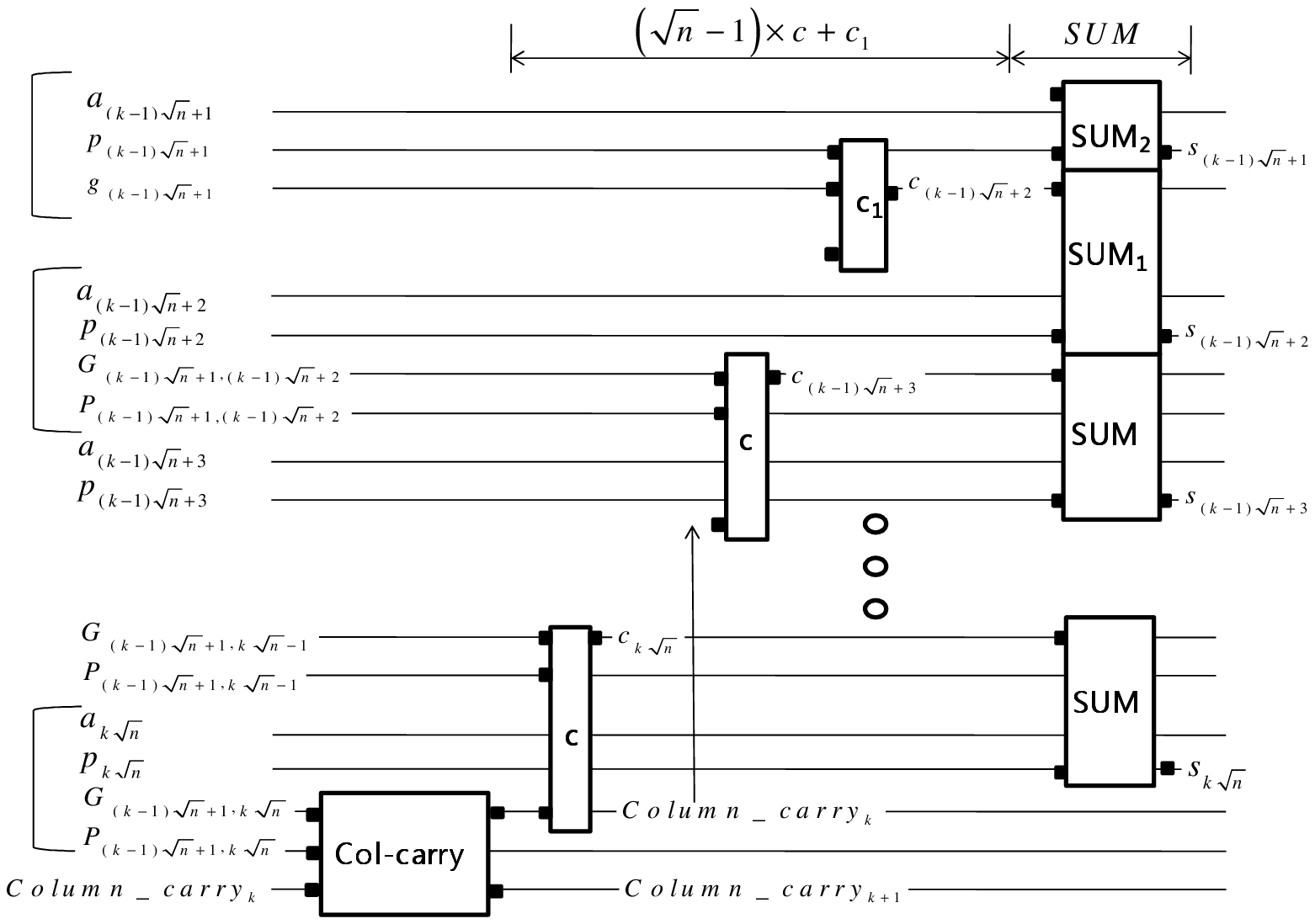,scale=0.8}} %100 percent
\vspace*{13pt}
\fcaption{\label{third-step-circuit-flow}Circuit flow for the third phase. \newline
\textbf{c} represents the block for generating carry output for $i$-th position. \textbf{SUM} is for generating the final summation value for $i$-th position.}
\end{figure}

Figure \ref{third-step-circuit-flow} shows the circuit blocks for the third phase. In the figure, \textbf{c} and \textbf{c1} represent the blocks for generating carry output for $i$-th position. Note for the first row, $p$ and $g$ are the same as $P$ and $G$, and hence the circuit block is slightly different. \textbf{SUM}, \textbf{SUM1}, and \textbf{SUM2} are for generating the final summation value for $j$-th position.

\subsection{Clearing Ancillae Qubits}

As shown in Table \ref{qubit-analysis}, three types of ancillae qubits are used, $c_i$, $P[i,j]$, and $Column\_carry_k$. To clean these ancillae, we have used the strategy proposed in Reference \cite{draper-QCLA}. The key idea of this approach is based on the observation that in two's complement arithmetic
\begin{eqnarray}
-x  &   \equiv  &   \bar{x} + 1 \pmod{2^n}, \\
\bar{x} + x &   \equiv  &    -1 \pmod{2^n}, \\
-x - 1      &   \equiv  &   \bar{x} \pmod{2^n},
\end{eqnarray}
where $\bar{x}$ is the bit-wise inversion of $x$.
Let us consider an addition of $A$ and $B$, $ADD(A,B,0)=(A,S,C)$, where $S$ and $C$ are the bitwise sum and carry vectors, respectively. Let us consider another addition of $A$ and $\bar{S}$, $ADD(A,\bar{S},0)=(A, \bar{B},D)$, where $\bar{B}$ and $D$ are sum and carry vectors, respectively. Note the bitwise sum is $\bar{B}$ because
\begin{equation}
A + \bar{S}    =    A-(A+B+1)     =    -B-1    =   \bar{B}.
\end{equation}
It is worth noting that $C$ must be equal to $D$ because of
\begin{eqnarray}
A \oplus \bar{S} \oplus D                                       &   =   &   \bar{B}, \\
A   \oplus  A   \oplus  B   \oplus  C   \oplus  1   \oplus  D   &   =   &   \bar{B}, \\
                        B   \oplus  C   \oplus  1   \oplus  D   &   =   &   \bar{B}, \\
                        \bar{B}   \oplus  C   \oplus  D         &   =   &   \bar{B}, \\
                                          C   \oplus  D         &   =   &   0,   \\
                                                    C           &   =   &   D.
\end{eqnarray}

\begin{figure}[t]
\centerline{\epsfig{file=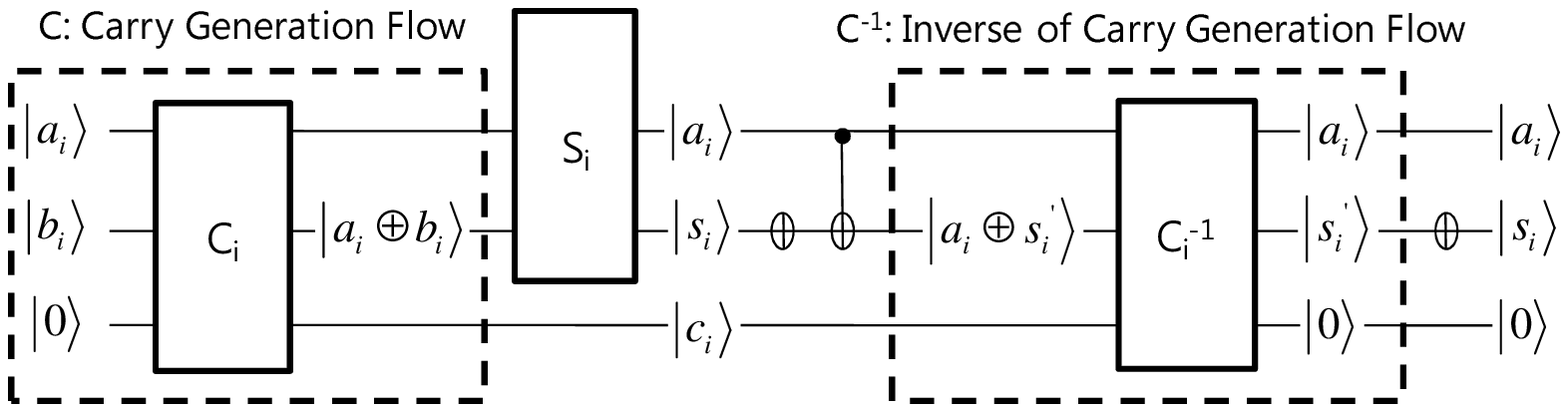,scale=0.6}} %100 percent
\vspace*{13pt}
\fcaption{\label{uncompute}Clearing Ancillae Qubits. \newline
By applying the inverse of the carry generation flow, the ancillae qubits can be cleaned.}
\end{figure}

Now we follow the circuit as shown in Figure \ref{uncompute}. Conceptually any addition circuit can be divided into two parts, $CARRY$ generation ($C_i$)and $SUM$ generation ($S_i$). As shown in the figure, we apply $CARRY$ as follows.
\begin{equation}
CARRY(A,B,0)    \Longrightarrow (A, A \oplus B, C).
\end{equation}
As the second step, we apply $SUM$ as follows.
\begin{equation}
SUM(A, A \oplus B, C)   \Longrightarrow (A, A \oplus B \oplus C, C) = (A, S, C).
\end{equation}
Apply two operations
\begin{equation}
NOT_2(A, S, C)  \Longrightarrow   (A,\bar{S}, C).
\end{equation}
\begin{equation}
CNOT_{1,2}(A, \bar{S}, C) \Longrightarrow (A, A \oplus \bar{S}, C).
\end{equation}
Meanwhile,
\begin{equation}
CARRY(A,\bar{S},0)  \Longrightarrow (A, A \oplus \bar{S}, D).
\end{equation}
Since the two carry vectors $C$ and $D$ for $A+B$ and $A+\bar{S}$ are the same, the above line changes to
\begin{equation}
CARRY(A,\bar{S},0)  \Longrightarrow (A, A \oplus \bar{S}, C).
\end{equation}
Therefore, running the inverse operation,
\begin{equation}
CARRY^{-1}(A, A \oplus \bar{S}, C) \Longrightarrow (A,\bar{S},0) .
\end{equation}
Finally, apply $NOT_2$ as follows.
\begin{equation}
NOT_2(A, \bar{S}, C) \Longrightarrow (A,S,0),
\end{equation}
to generate the final sum and clean ancillae.

\section{Analysis}
\label{sec:analysis}

\subsection{Depth Analysis}

\begin{figure}[t]
\centerline{\epsfig{file=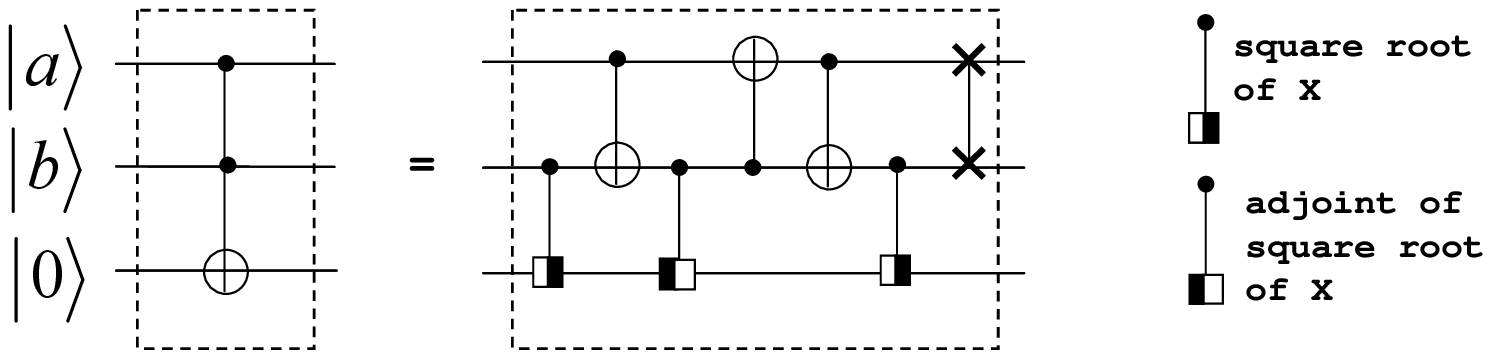,scale=0.40}}
\vspace*{13pt}
\fcaption{\label{ccnot}Circuit for \textbf{CCNOT}}
\end{figure}

To analyze the depth of the proposed adder, we have to decompose the circuit blocks into elementary gates, which can be decomposed into unit delay gates. In this work, we assume one-qubit, \textbf{CNOT}, and \textbf{Control-$\sqrt{NOT}$} gates have unit delay. The elementary gates we have chosen for constructing our circuits are \textbf{SWAP}, \textbf{CCNOT}, \textbf{CNOT}, \textbf{Control-$\sqrt{NOT}$}, and one-qubit gates. In this paper, we use the three-\textbf{CNOT} construction for \textbf{SWAP} gate. Figure \ref{ccnot} shows the conventional form of \textbf{CCNOT} (left) and its decomposition into one-qubit and two-qubit gates (right).

\subsubsection{Circuit Decomposition with NTC Constraints}
\label{Circuits-for-Each-Block-with-NTC-Condition}
\begin{figure}[t]
\centering
\subfigure[\label{half-adder}]{\epsfig{file=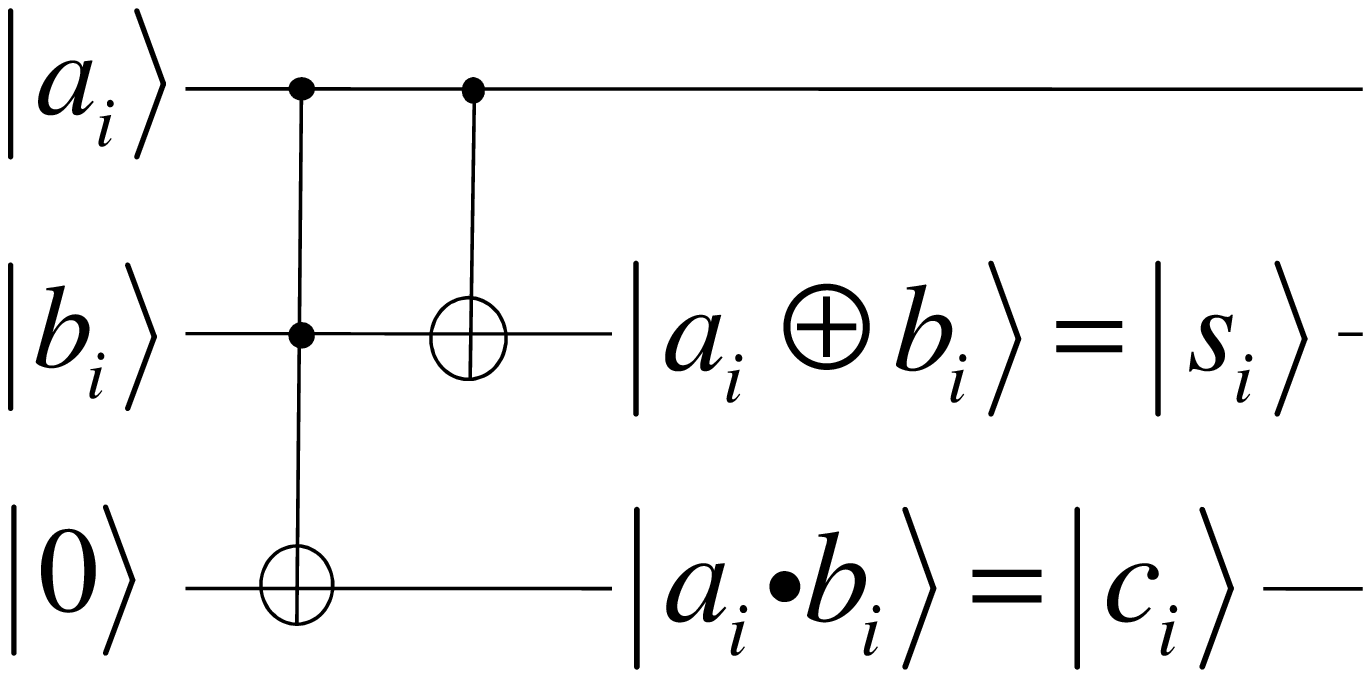,scale=0.20}} \newline
\subfigure[\label{full-adder}]{\epsfig{file=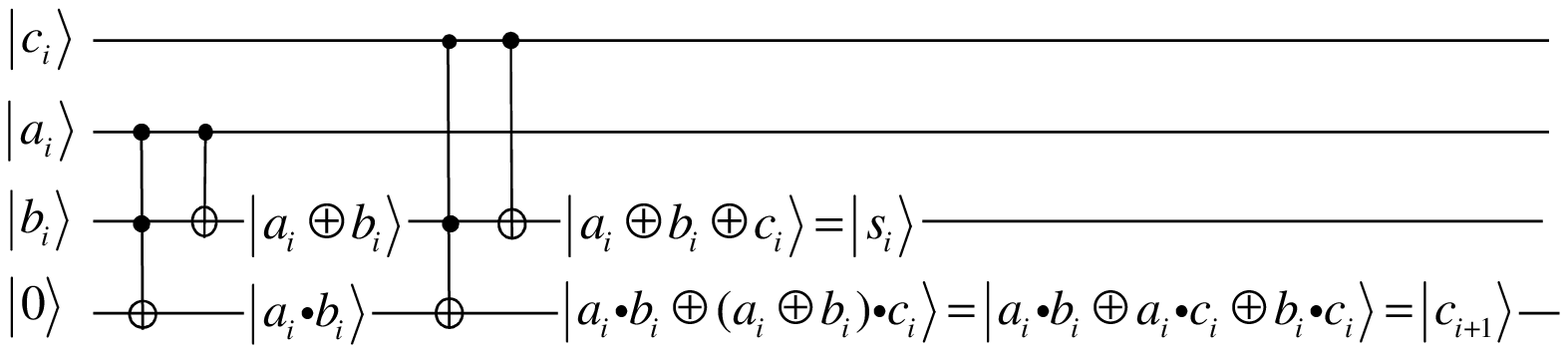,scale=0.55}}
\hspace*{10pt}
\subfigure[\label{full-adder-with-neighbor-interaction-only}]{\epsfig{file=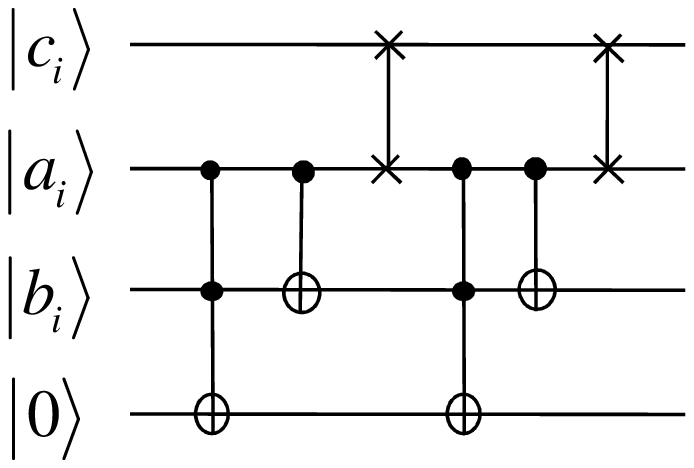,scale=0.40}}
\fcaption{\label{figure:Adders}Circuit for \textbf{HALF ADDER} (a); Circuits for \textbf{FULL ADDER} with arbitrary interaction (b) and with only nearest-neighbor interaction (c)}
\end{figure}

\begin{figure}[t]
\centering
\subfigure[\label{smallgp}]{\epsfig{file=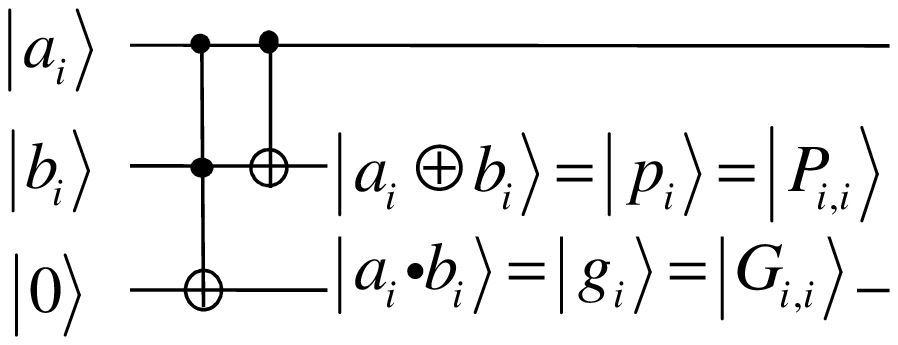,scale=0.40}} \newline
\subfigure[\label{largeGP}]{\epsfig{file=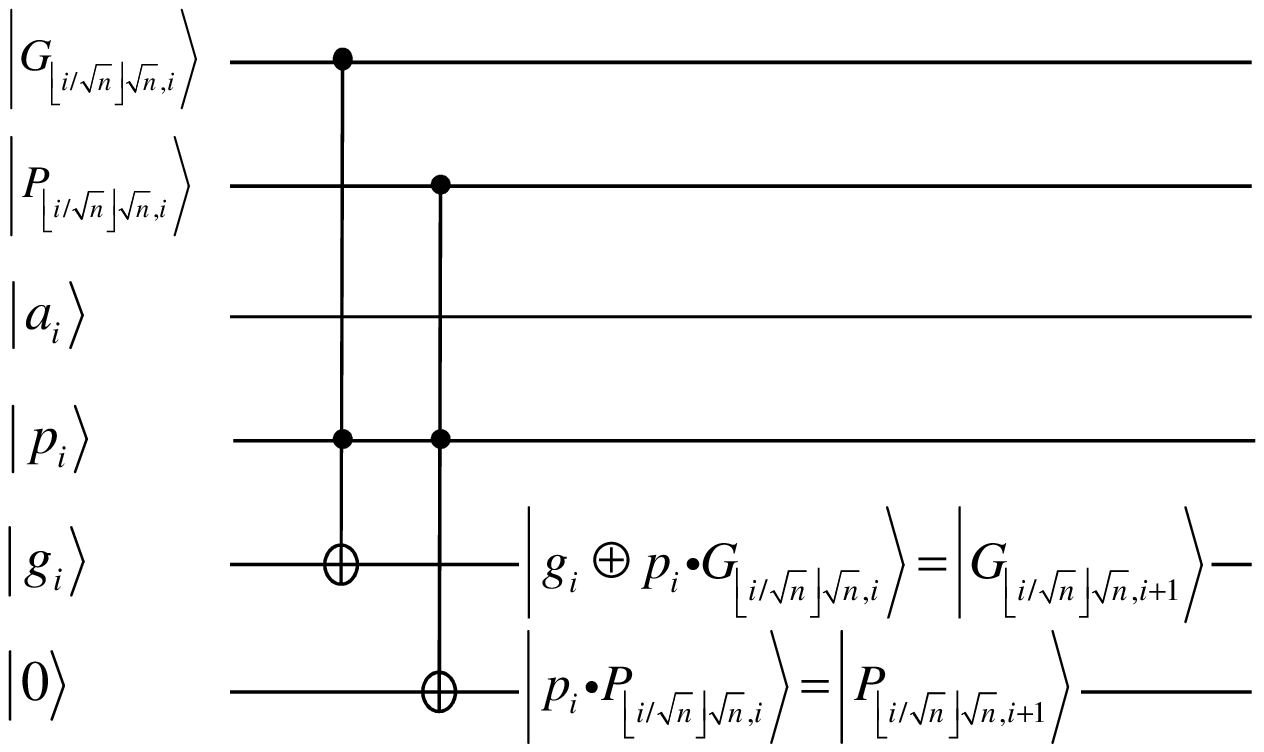,scale=0.4}}
\hspace*{10pt}
\subfigure[\label{largeGP-with-neighbor-interaction-only}]{\epsfig{file=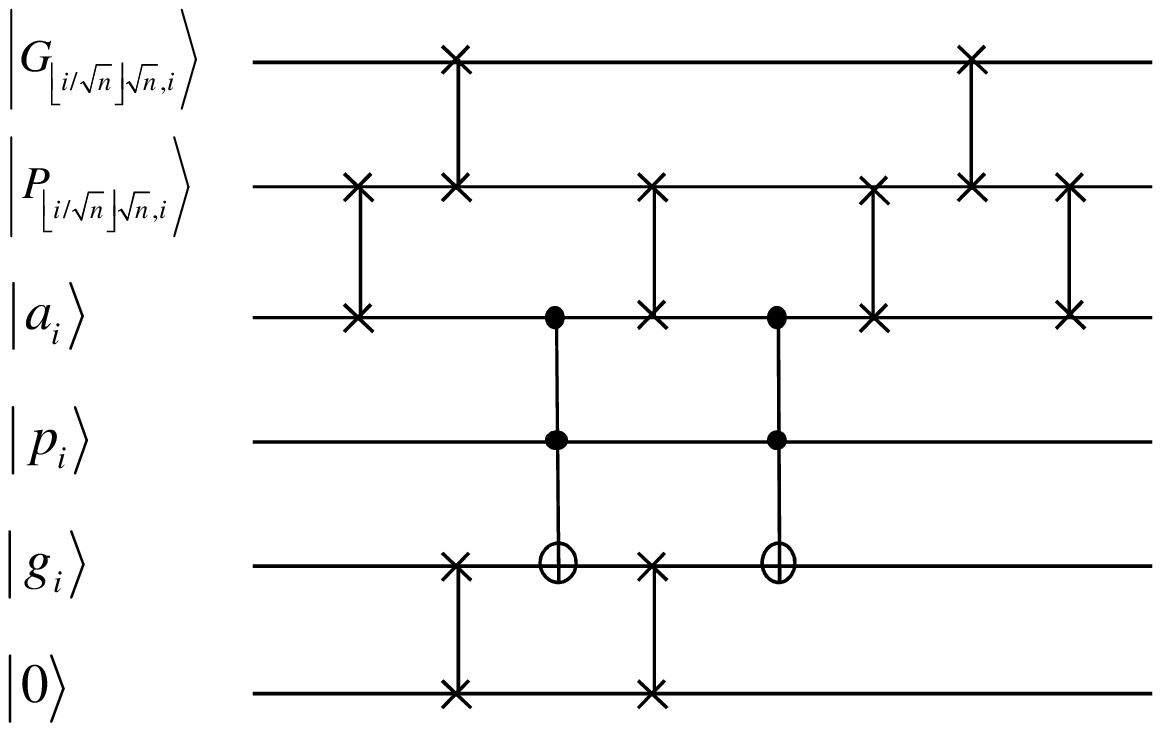,scale=0.4}}
\fcaption{\label{figure:largeGP}Circuits for \textbf{g,p} (a); Circuits for \textbf{G} and \textbf{P} with arbitrary interaction (b) and with only nearest-neighbor interaction (c)}
\end{figure}

\begin{figure}[t]
\centering
\subfigure[\label{column_carry}]{\epsfig{file=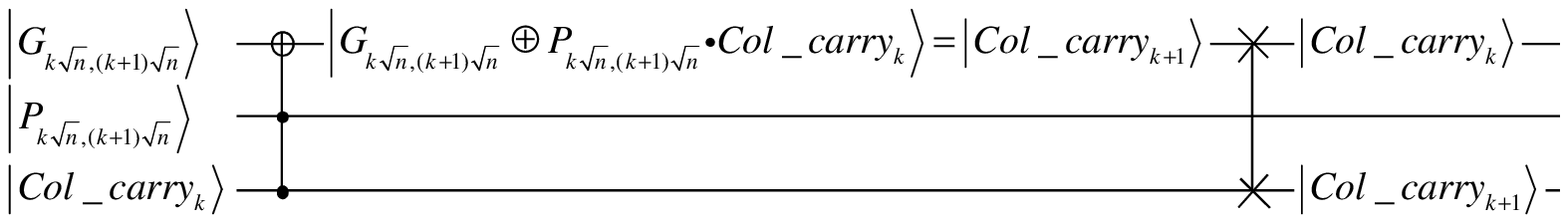,scale=0.55}}
\hspace*{10pt}
\subfigure[\label{column_carry-with-neighbor-interaction-only}]{\epsfig{file=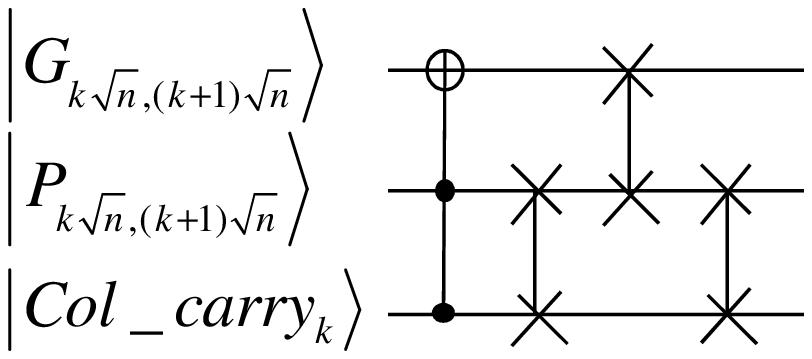,scale=0.33}}
\fcaption{\label{figure:column_carry}Circuits for \textbf{Column\_carry} with arbitrary interaction (a) and with only nearest-neighbor interaction (b)}
\end{figure}

\begin{figure}[t]
\centering
\subfigure[\label{carry}]{\epsfig{file=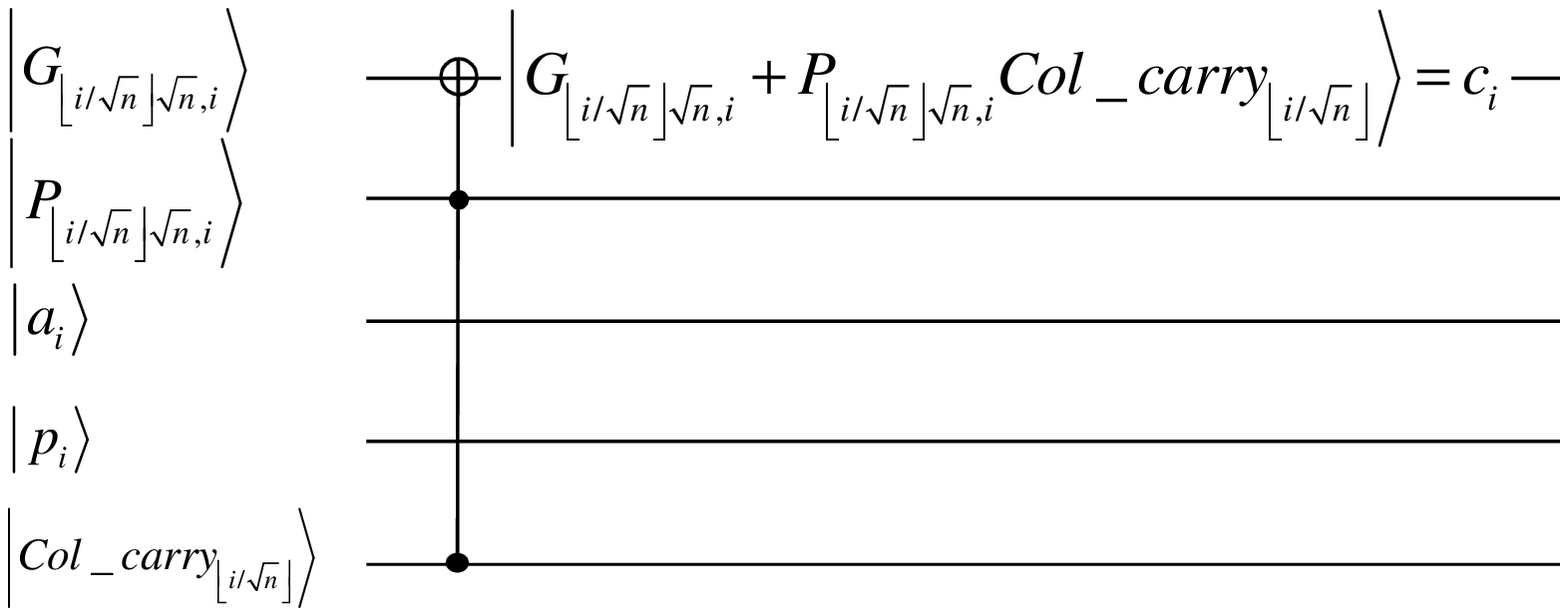,scale=0.4}}
\hspace*{10pt}
\subfigure[\label{carry-with-neighbor-interaction-only}]{\epsfig{file=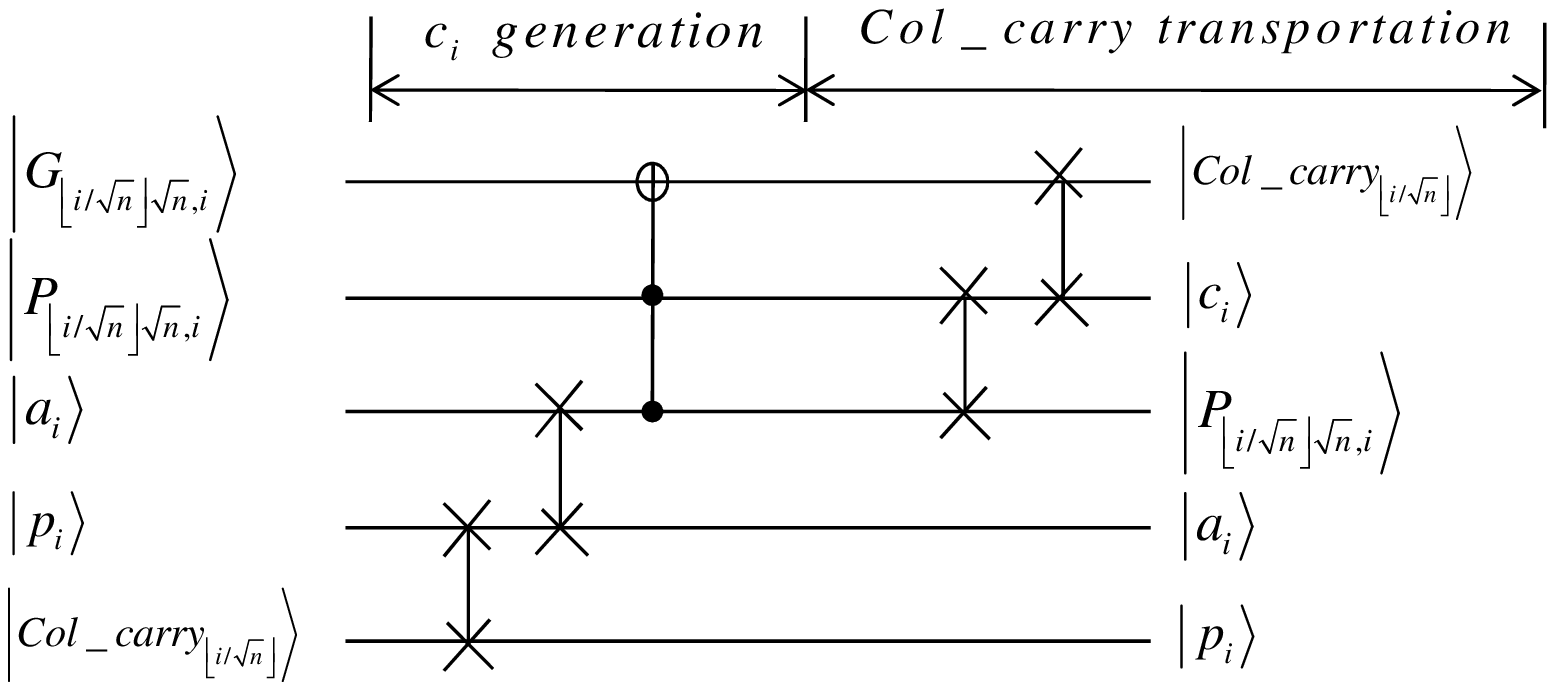,scale=0.4}}
\hspace*{10pt}
\subfigure[\label{carry1}]{\epsfig{file=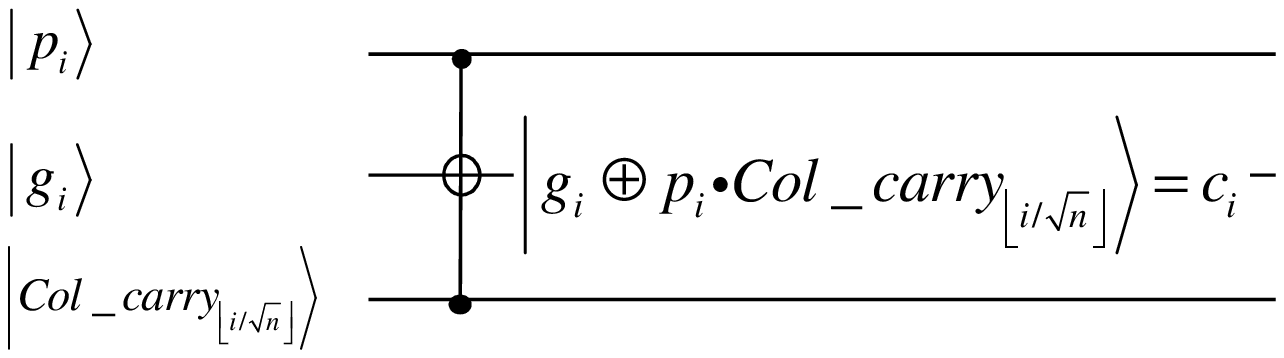,scale=0.4}}
\hspace*{10pt}
\subfigure[\label{carry1-with-neighbor-interaction-only}]{\epsfig{file=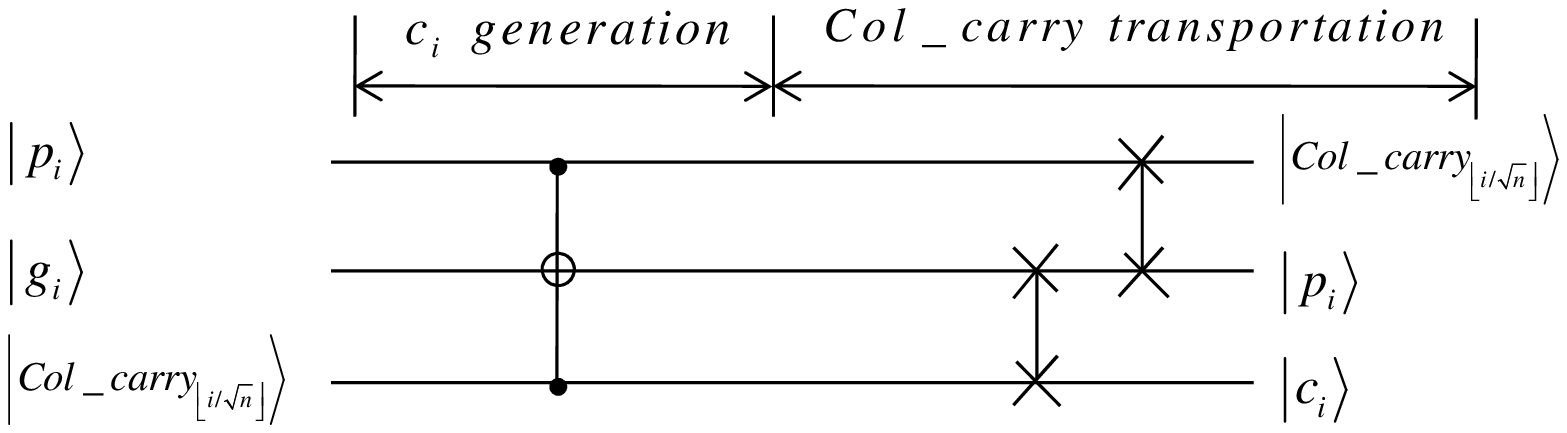,scale=0.44}}
\fcaption{\label{figure:carry}
Circuits for \textbf{Carry} with arbitrary interaction (a) and with only nearest-neighbor interaction (b); Circuits for \textbf{Carry1} with arbitrary interaction (c) and with only nearest-neighbor interaction (d)}
\end{figure}

\begin{figure}[t]
\centering
\subfigure[\label{sum}]{\epsfig{file=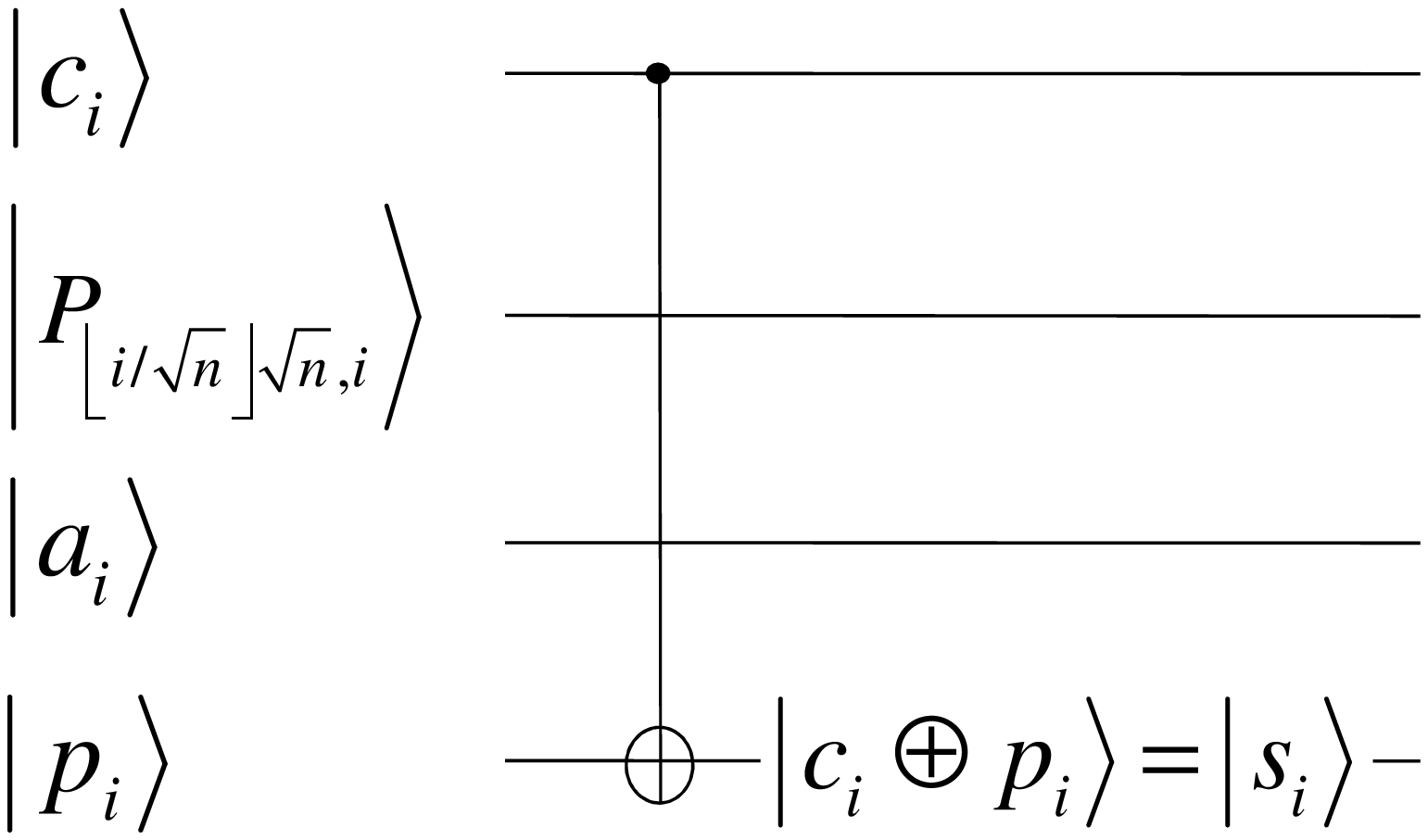,scale=0.2}}
\hspace*{10pt}
\subfigure[\label{sum-with-neighbor-interaction-only}]{\epsfig{file=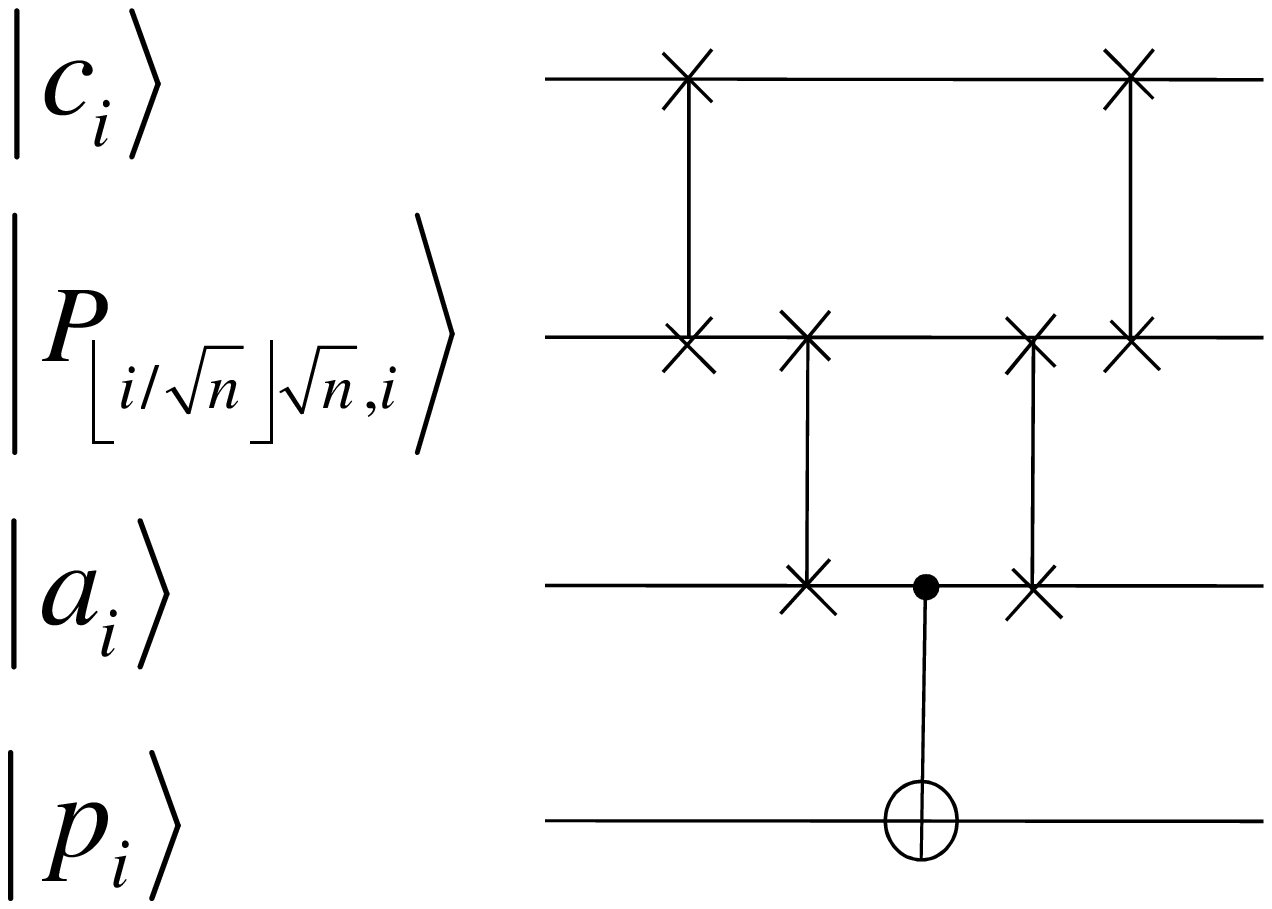,scale=0.2}}
\hspace*{10pt}
\subfigure[\label{sum1}]{\epsfig{file=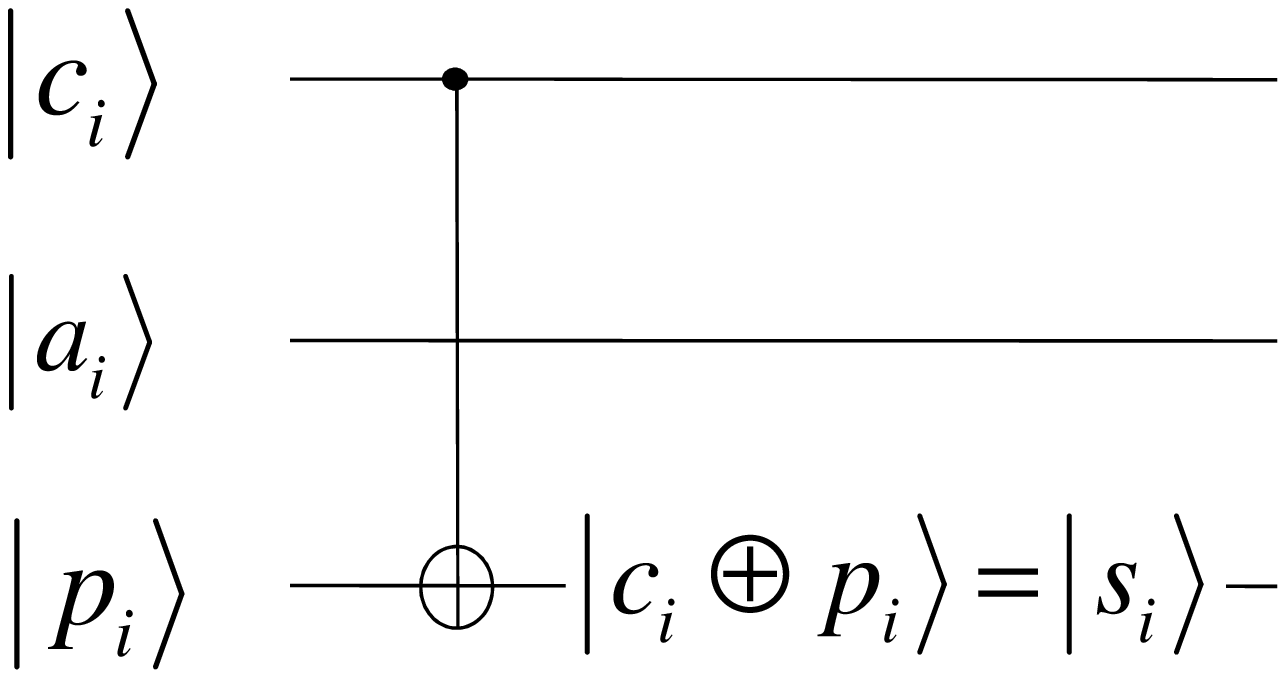,scale=0.2}}
\hspace*{10pt}
\subfigure[\label{sum1-with-neighbor-interaction-only}]{\epsfig{file=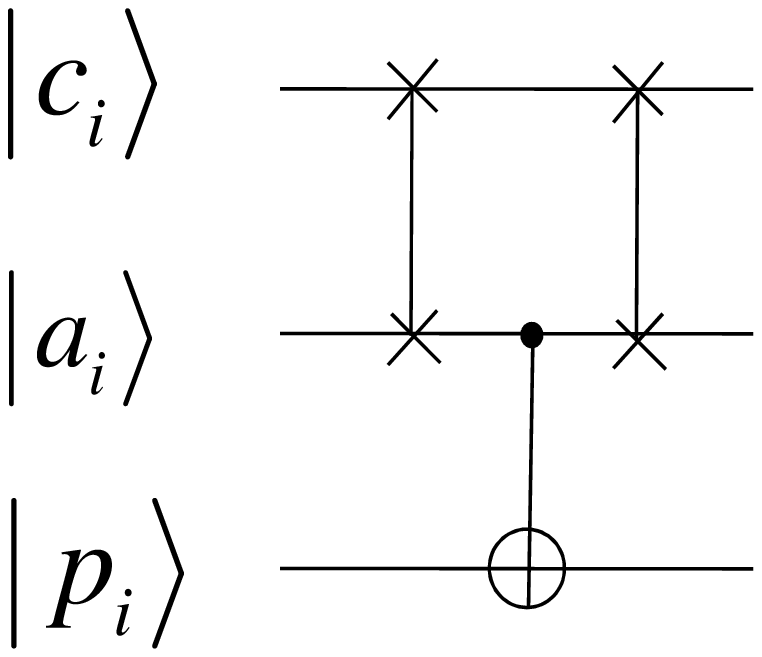,scale=0.2}}
\hspace*{10pt}
\subfigure[\label{sum2}]{\epsfig{file=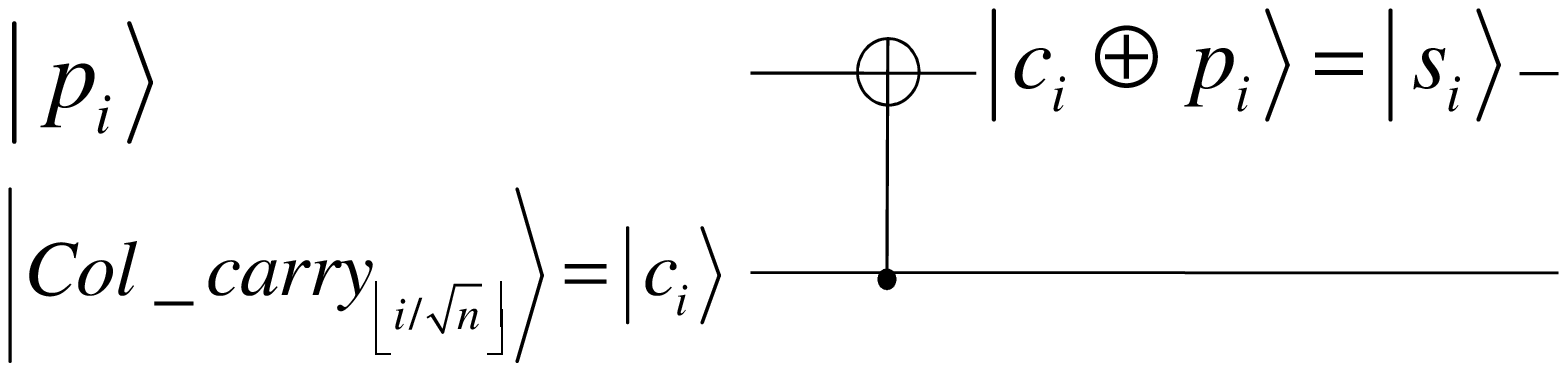,scale=0.225}}
\fcaption{\label{figure:sum}
Circuits for \textbf{SUM} with arbitrary interaction (a) and with only nearest-neighbor interaction (b); Circuits for \textbf{SUM1} with arbitrary interaction (c) and with only nearest-neighbor interaction (d); Circuit for \textbf{SUM2} (e)}
\end{figure}

Now we decompose the circuit blocks for three phases with the chosen elementary gates and necessary \textbf{SWAP} operations to satisfy the NTC constraints. The blocks are shown in Figures \ref{figure:Adders} to \ref{figure:sum}. The circuit of \textbf{HALF ADDER} is shown in Figure \ref{half-adder}. Figure \ref{full-adder} \cite{Cheng-Tseng-adder} shows a decomposition of \textbf{FULL ADDER} into elementary gates. In this figure, there is no limitation on the distance between operands for a gate. To satisfy the NTC constraint, we redesign it as shown in Figure \ref{full-adder-with-neighbor-interaction-only} by adding several \textbf{SWAP} gates to move the qubits to neighboring positions. This approach is also applied for the following circuit blocks. The circuits for \textbf{g} and \textbf{p}, and the generalized \textbf{G} and \textbf{P} are shown in Figure \ref{figure:largeGP}. The circuit of \textbf{Column\_carry} is shown in Figure \ref{figure:column_carry}. For generating $|Col\_Carry_{k+1}\rangle$, a single \textbf{CCNOT} is enough. However, to propagate it to the next column and to propagate $|Col\_Carry_{k}\rangle$ to the rows, a \textbf{SWAP} is necessary. For implementing the last \textbf{SWAP} gate in the neighbor interaction only case, several \textbf{SWAP}s are necessary as shown in Figure \ref{column_carry-with-neighbor-interaction-only}. The initial circuit for \textbf{Carry} is shown in Figure \ref{carry}. Since the $Col\_carry$ has to be moved to the upper row, several \textbf{SWAP}s are necessary as shown in Figure \ref{carry-with-neighbor-interaction-only}. After this circuit, the $Col\_carry$ is transported to the top position, and the others are to the lower row. Since the carry for the first row is different from other rows, Figures \ref{carry1} and \ref{carry1-with-neighbor-interaction-only} show its circuits. The circuits for \textbf{SUM} are shown in Figures \ref{sum} and \ref{sum-with-neighbor-interaction-only}. For the second and the first row, we have to use slightly different circuits as shown in Figures \ref{sum1} and \ref{sum1-with-neighbor-interaction-only}, and Figure \ref{sum2}, respectively.

\subsubsection{Total Depth}

Based on the revised circuits with satisfying the NTC constraint, we can summarize the depth of each elementary gate and circuit block as shown in Table \ref{depth-analysis}.

\vspace*{4pt}
\begin{table}[t]
\tcaption{\label{depth-analysis}Depth analysis of each gate and circuit}
\centerline{\footnotesize\smalllineskip
\begin{tabular}{l|l|r}  \hline
Name & Composition of the longest path & \# of unit-gate steps                  \\ \hline   \hline
  \textbf{SWAP}             &   3 CNOTs                                    & 3 \\
  \textbf{CCNOT}            & 1 SWAP + 6 unit gates                        & 9  \\
  \textbf{HALF ADDER}       & 1 CCNOT + 1 CNOT                             & 10 \\
  \textbf{FULL ADDER}       & 2 CCNOTs + 2 CNOTs + 2 SWAPs                 & 26 \\
  \textbf{g} and \textbf{p} & 1 CCNOT + 1 CNOT                             & 10 \\
  \textbf{G} and \textbf{P} & 2 CCNOTs + 6 SWAPs                           & 36 \\
  \textbf{Column\_carry}    & 1 CCNOT + 3 SWAPs                            & 18 \\
  \textbf{Carry}            & 1 CCNOT + 4 SWAPs                            & 21 \\
  \textbf{Carry1}          & 1 CCNOT + 2 SWAPs                            & 15 \\
  \textbf{SUM}              & 1 CNOT  + 4 SWAPs                            & 13 \\
  \textbf{SUM1}            & 1 CNOT  + 2 SWAPs                            & 7  \\
  \textbf{SUM2}            & 1 CNOT                                       & 1  \\ \hline
\end{tabular}}
\end{table}

The proposed adder works in three sequential phases, and hence the overall depth is the sum of the depths for each phase. The depth for each phase is the ``long pole", or the longest delay among the parallel execution  paths. In the first column, one \textbf{HA} and $(\sqrt{n}-1)$ \textbf{FA} operations are executed sequentially. Since \textbf{HA} needs 10 unit-gate steps and \textbf{FA} needs 26 unit-gate steps, $26\sqrt{n}-16$ unit-gate steps are needed. On the other hand, the other columns need one \textbf{g,\,p} + ($\sqrt{n}$-1)\textbf{G,\,P}, which is $36\sqrt{n}-26$. The overall depth for the first phase is the longer of the two column types, hence $36\sqrt{n}-26$. The second phase consists of $(\sqrt{n}-1)$ $\textbf{Column\_carry}$ operations, requiring a total of $18\sqrt{n}-18$ time steps. The third phase consists of $(\sqrt{n}-1)$ \textbf{Carry} + \textbf{Carry1} and \textbf{SUM1} operations for the longest path. Hence, the depth is $21\sqrt{n}+1$ unit-gate steps. By summing depths of each phase, the total depth is $75\sqrt{n}-43$.

The above depth is only for generating the summation output without clearing the ancillae. For clearing ancillae, we apply more circuits as shown in Figure \ref{uncompute}. Based on this figure, we can decompose the above three phases into the carry generation flow and the sum generation flow. The first and the second phases are for the carry generation flow. The third phase has to be divided into the carry generation flow and the sum generation flow. The above depth is apportioned as $75\sqrt{n}-50$ for carry generation flow and $7$ for sum generation flow. As shown in Figure \ref{uncompute} we need to apply \textbf{NOT} and \textbf{CNOT} gates and then inverse of the carry generation flow again with the final \textbf{NOT} gate. Hence, the overall depth is $75\sqrt{n}-50 + 7 + 1 + 1 + 75\sqrt{n}-50 + 1 = 150\sqrt{n}-90$.

\subsection{Required Space}

The number of qubits for the adder is shown in Table \ref{qubit-analysis}. As shown in the first column, some qubits work for multiple purposes. Note the additional number of qubits is $2n-\sqrt{n}$, which is less than twice the minimum $2n$ qubits \cite{CDKM-adder,takahashi-2005}.

\vspace*{4pt}
\begin{table}[t]
\tcaption{\label{qubit-analysis}Number of Qubits}
\centerline{\footnotesize\smalllineskip
\begin{tabular}
{l|l|p{280pt}}  \hline
Name & Number of qubits & Explanation \\ \hline   \hline
$a_i$ & $n$ & Input $A$ \\ \hline
$b_i$ $\rightarrow$ $p_i$ $\rightarrow$ $s_i$ & $n$ & Input $B$, Carry propagate for $i$-th position, and Summation $S$\\ \hline
$|0\rangle$ $\rightarrow$ $g_i$ $\rightarrow$ $G[i,j]$ $\rightarrow$ $c_i$ & $n$ & Carry generation for $i$-th position, Carry generation between $i$ and $j$, and carry for $i$-th position\\ \hline
$|0\rangle$ $\rightarrow$ $P[i,j]$ & $ n-2\sqrt{n}+1$ & Carry propagation between $i$ and $j$ \\  \hline
$Column\_carry_k$ & $\sqrt{n}$ & Inter column carry. The last $Column\_carry$ is for the final carry output. \\ \hline  \hline
Total & ($2n+1$)+($2n-\sqrt{n}$) & Mandatory + Additional \\ \hline \hline
\end{tabular}}
\end{table}

\subsection{Comparison to Other Adders}

\vspace*{4pt}
\begin{table}[t]
\tcaption{\label{table:comparison}Comparison with Other Designs}
\centerline{\footnotesize\smalllineskip
\begin{tabular}
{l|l|l|p{80pt}|l} \hline
Architecture & Name of Adder                               & (Depth, Number of Qubits)                        & When is the present adder faster than the corresponding adder? & KQ\cite{KQ-analysis} \\ \hline \hline
\raisebox{-3.25ex}[0cm][0cm]{1D NTC}       & VBE\cite{VBE-adder}                         &  $(76n-30,\,\, 3n+1)$ &                                $n \geq 4$       & $228n^2-O(n)$ \\
\cline{2-5}             & VBE-Improved\cite{VBE-Improved}             &  $(20n-15,\,\, 3n+1)$                                & $n \geq 49 $      & $60n^2-O(n)$ \\
\cline{2-5}             & CDKM\cite{CDKM-adder}                       &  $(18n+14,\,\, 2n+2)$                                & $n \geq 58$      & $36n^2+O(n)$ \\ \hline
2D NTC       & Present Adder                               &  ($150\sqrt{n}-90$,\,\,$4n-\sqrt{n}+1$)              &            & $600 n\sqrt{n}-O(n)$ \\ \hline \hline
\raisebox{-3.25ex}[0cm][0cm]{AC}          & QFT-based\cite{Draper-QAddition}            & ($3\log{n}$,\,\, $2n+1$)                         &  N/A & $6n\log{n} +O(\log{n})$  \\
\cline{2-5}             & CLA-based\cite{draper-QCLA}                 & ($2\log{n}+2$,\,\, $4n-\log{n}$)                 &    N/A      & $8n\log{n}+O(n)$   \\
\cline{2-5}             & RCA+CLA-based\cite{KAWATAYoshinori:20080229}& ($10\log{n}+6n/{\log{n}}$,\,\, $n+4n/{\log{n}}$) & N/A  & $10n\log{n}+O(n^2/ \log{n})$  \\ \hline
\end{tabular}}
\end{table}

When only interactions between neighboring qubits are allowed, the depth of arithmetic circuits increases. For the 2D case, the depth lower bound was proven to be $\Omega(\sqrt{n})$ \cite{choi-2008}. Therefore, the depth of the proposed adder is asymptotically optimal.

Beyond the asymptotic behavior, it seems more interesting and important to compare with other adders in the practical cases. Specifically, it is necessary to compare adders designed for the 1D NTC architecture since they can be implemented on the 2D NTC architecture without modification, using a simple serpentine qubit layout. The overall analysis and the comparison between the adders are shown in Table \ref{table:comparison}. The first column distinguishes the architecture and the second column lists the adder type. For the 1D NTC architecture, we choose three typical adders. Vedral et al. proposed a plain ripple-carry adder \cite{VBE-adder}, named \emph{VBE} in the table. \emph{VBE-Improved} is the Van Meter and Itoh update to this adder \cite{VBE-Improved}. Cuccaro et al. proposed a ripple carry adder with only one ancillae qubit \cite{CDKM-adder}, named \emph{CDKM}. For the 2D NTC architecture, the present adder is shown. For the architecture with arbitrary distance interaction, several adders are evaluated. Draper proposed a quantum Fourier transform adder \cite{Draper-QAddition}, named \emph{QFT-based}. By exploiting the classical fast addition algorithm, Draper et al. also proposed a carry-lookahead adder \cite{draper-QCLA}, named \emph{CLA-based}. Kawata et al. also proposed an adder based on the combination of ripple carry adder and carry-lookahead adder \cite{KAWATAYoshinori:20080229}, named \emph{RCA+CLA-based}. For comparison, the depth and the size of each adder is shown in the third column. In this work, the depth is measured by in units of one- and two-qubit gates for the 1D and 2D NTC architectures. On the other hand, the depth for the \textbf{AC} architecture is based on one-, two-, and CCNOT gates. The size is for the number of qubits for input, output, and ancillae. In the fourth column, the input size is shown when the selected adder works faster than the present adder. In the fifth column, we calculate \emph{KQ}, the product of qubits and depth where $K$ and $Q$ are the numbers of logical qubits and computational steps, respectively \cite{KQ-analysis}. KQ is used to estimate the strength of error correction required.

From this table we can point out three key results. First, when the size of input is larger than 58, the present adder works faster than 1D NTC adders. Second, the present adder needs about two times number of qubits than 1D NTC adders. Lastly, the present adder has a smaller KQ factor when the input size is larger than 278.

\section{Conclusion and Open Problems}
\label{sec:conclusion}

In this work, we proposed a quantum adder for the 2D NTC architecture for the first time. Van Meter and Oskin indicated that an adder would be in $O(\sqrt{n})$ time complexity on a 2D architecture, but no circuit has been provided \cite{VanMeter-Oskin-2006}. The proposed adder has the depth complexity $\Theta(\sqrt{n})$ with $O(n)$ qubits. We found that the proposed adder works faster than a 1D ripple-carry adder when the length of the input registers is larger than 58, and requires about two times the number of additional qubits.

Although this adder is, to the best of our knowledge, the first one specifically designed for a 2D architecture, we suspect it will not be the last; we anticipate that several improvements are possible. First, the number of additional gates is very large. Most of the gates for the proposed adder are used for transporting qubits to neighboring positions so that gates can be executed. By arranging qubits in a better way, we may be able to reduce the necessary propagattion operations. Second, the phase for cleaning the ancillae qubits roughly doubles the total number of quantum operations. In the present adder, the ancillae qubits are reinitialized by applying the inverse circuit, doubling the overall depth. Perhaps there is some way to reduce this drawback by exploiting some overlap of the clearing phase with the computation phase. Third, the number of ancillae is also very large. The proposed design attempts to achieve the highest parallel execution at the expense of requiring more ancillae, but this tradeoff may prove to be less than optimal for two reasons. First, qubits themselves are expensive resources, and in many applications could be allocated to other work if not used directly in the adder; second, inserting the ancillae into our layout increases the distance between qubits, forcing the addition of more \textbf{SWAP}s and slowing down the circuit.

\nonumsection{Acknowledgements}
\noindent
This research is supported in part by the Japan Society for the Promotion of Science (JSPS) through its ¡°Funding
Program for World-Leading Innovative R\&D on Science and Technology (FIRST Program),¡± and in part by the National Research Foundation of Korea Grant funded by the Korean Government(Ministry of Education, Science and Technology).[NRF-2010-359-D00012]

\nonumsection{References}

\bibliographystyle{unsrt}
\bibliography{../../../../../../../../reference-list}
\end{document}